\documentclass[twocolumn]{aastex63}

\usepackage{diagbox}
\usepackage{amsmath}
\usepackage{graphicx}
\usepackage{subfigure}
\received{xxx, xxx}
\revised{xxx, xxx}
\accepted{xxx, xxx}

\shortauthors{Zhang \& Guo}
\defcitealias{Zhang2020}{Z20}

\begin{document}

\title{Probing the Halo Gas Distribution in the Inner Galaxy with Fermi Bubble Observations}
\shorttitle{Probing the halo gas distribution with Fermi bubble observations}
\author{Ruiyu Zhang}
\affiliation{Key Laboratory for Research in Galaxies and Cosmology, Shanghai Astronomical Observatory, Chinese Academy of Sciences,
 80 Nandan Road, Shanghai 200030, China }
\affiliation{University of Chinese Academy of Sciences, 19A Yuquan Road, Beijing 100049, China}

\author{Fulai Guo}
\affiliation{Key Laboratory for Research in Galaxies and Cosmology, Shanghai Astronomical Observatory, Chinese Academy of Sciences, 80 Nandan Road, Shanghai 200030, China }
\affiliation{University of Chinese Academy of Sciences, 19A Yuquan Road, Beijing 100049, China}

\correspondingauthor{Fulai Guo}
\email{fulai@shao.ac.cn}

\begin{abstract}

The hot halo gas distribution in the inner Milky Way (MW) contains key fossil records of the past energetic feedback processes in the Galactic center. Here we adopt a variety of spherical and disk-like MW halo gas models as initial conditions in a series of simulations to investigate the formation of the Fermi bubbles in the jet-shock scenario. The simulation results are compared directly with relevant X-ray and gamma-ray observations of the Fermi bubbles to constrain the halo gas distribution in the inner Galaxy before the Fermi bubble event. Our best-fit gas density distribution can be described by a power law in radius $n_{\rm e}(r)=0.01(r/1 \text{~kpc})^{-1.5}$ cm$^{-3}$. Our study can not determine if there is an inner density core, which if exists, should be very small with size $r_{c} \lesssim 0.5$ kpc. When extrapolating to large radii $r\sim 50-90$ kpc, our derived density distribution lies appreciably below the recently estimated gas densities from ram-pressure stripping calculations, suggesting that the halo gas density profile either flattens out or has one or more discontinuities within $10 \lesssim r \lesssim 50$ kpc. { Some of these discontinuities may be related to the eROSITA bubbles, and our derived gas density profile may correspond to the hot gas distribution in the inner eROSITA bubbles about $5$ Myr ago.}

\end{abstract}

\keywords{Circumgalactic medium -- Cosmic rays -- Galaxies: jets -- Galaxy : halo -- Gamma rays: galaxies -- Methods: numerical -- Milky Way Galaxy -- X-rays: galaxies }

\section{Introduction}
\label{sec:intro}

Galaxy formation theories predict that, surrounding a Milky-Way-like galaxy, there exists a hot gaseous halo formed via cosmic accretion of the intergalactic medium with the gas heated to around the virial temperature \citep{White1978,White1991,Kerevs2005}. The gaseous halo, usually referred to as the circumgalactic medium (CGM), provides fuel for star formation activities in the central galaxy, which in turn inject mass, momentum and energy back to the gaseous halo through feedback processes \citep[see][]{Nuza2014,Li2020,Martizzi2020}. The baryonic mass stored in the gaseous halo may also help explain the cosmological ''missing baryon'' problem \citep{Fukugita2006, Fang2013}. Owing to the essential role that the gaseous halo plays in galaxy formation and evolution, significant efforts have been made over the past several decades in this field \citep[for a comprehensive review, see][and references therein]{Putman2012,Tumlinson2017}.

For a galaxy with the virial mass $M_{\rm vir}$, the virial temperature of the CGM at the virial radius is $T_{\rm vir}\sim 5\times 10^{5}(M_{\rm vir}/10^{12}M_{\sun})^{2/3}$ K, where $M_{\sun}$ is the solar mass. At smaller radii, the gas temperature is expected to further rise due to adiabatic compression and galactic feedback processes. For a Milky-Way-like galaxy, the hot CGM radiates significantly in soft X rays, and diffuse X-ray observations have been extensively studied for nearby galaxies \citep[e.g.][]{Forman1985,Wang2001,Strickland2004,Tyler2004,Anderson2011} and for the Milky Way \citep[MW;][]{Tanaka1977, Snowden1997, Henley2013, Nakashima2018, Nakahira2020, Kaaret2020, predehl20}. The X-ray-measured CGM temperature has even been recently used to constrain the MW mass $M_{\rm vir}$ and the measured MW mass lies on the high mass side of current $M_{\rm vir}$ estimates derived from various traditional kinematic methods \citep{Guo2020}.

Due to its proximity, the MW CGM is of particular importance and has been extensively studied with a variety of additional methods. The earliest evidence for an extended MW CGM came from observations of HI clouds in the Galactic halo, which may be in pressure equilibrium with an ambient hot diffuse medium \citep{Spitze1956}. Further pieces of evidence include emission and absorption line observations of O VII and O VIII at redshift $z\sim 0$ \citep[e.g.][]{Nicastro2002,Williams2005,Gupta2012,Miller2013,Miller2015,Li2017}, dispersion measure measurements of distant pulsars \citep{McQuinn2014,Nugaev2015,Keating2020}, and the ram-pressure stripping arguments of the MW satellite galaxies \citep{Grcevich2009,Gatto2013,Salem2015}.

Although the MW CGM has been studied for over half a century, the spatial distributions of its metallicity, density, and temperature are still poorly understood. We are located near the center of the MW halo, and it is difficult to derive these spatial distributions from line-of-sight-integrated X-ray emissions alone. While investigating the MW CGM with X-ray observations, one often assumes that the gas temperature and metallicity are spatially uniform, and further adopts an analytic model for the CGM density distribution. The proposed density models can be divided into two categories: spherical models and disk-like models. If cosmic inflows dominate the MW CGM formation, a spherical gaseous halo is expected to form \citep[see:][]{Cavaliere1976,Maller2004,Mathews2017,Sormani2018,Voit2019,Guo2020}. On the other hand, it was also argued that the halo gas distribution would be disk-like if the hot CGM is mainly contributed by stellar feedback processes in the Galactic disk \citep[e.g.,][]{Yao2005,Sofue2019}. As the latter could not explain the halo gas densities constrained by the ram-pressure stripping arguments at large distances, a hybrid density model with an inner disk-like distribution and an outer spherical distribution has also been proposed \citep{Kaaret2020}.

In this paper, we focus on the halo gas distribution in the inner Galaxy ($r<10$ kpc), where active galactic nucleus (AGN) feedback and stellar feedback from the Galactic bulge interact most strongly with the MW CGM. When using different CGM density models to fit X-ray observations, the derived gas density distribution often varies significantly, and in particular, the gas densities in the inner Galaxy could vary by several orders of magnitude (see Fig. \ref{fig-den1d}). The ram-pressure stripping method has been used to determine the gas densities at the locations of a few MW satellites (see Table 7 in \citealt{Bland-Hawthorn2016}). However, these satellites are all located at large distances, and could not probe the halo gas distribution in the inner Galaxy. Therefore, studies of the halo gas density distribution in the inner Galaxy may provide a key constraint on the MW CGM density model, or perhaps more interestingly, help reveal the physics of feedback processes at the Galactic center (GC).

The halo gas in the inner Galaxy could be strongly disturbed by AGN and stellar feedback processes originating at the GC. The most notable diffuse structure in the inner Galaxy may be the Fermi bubbles, which are two giant gamma-ray bubbles located symmetrically above and below the Galactic plane \citep{Dobler2010,Su2010,Ackermann2014}. Numerous recent studies strongly suggest that the Fermi bubbles are a Galactic scale phenomenon associated with energetic outbursts from the GC \citep[e.g.,][]{Crocker2011,Guo2012,Guo2012a,Yang2012,Fujita2013,Lacki2014,Mou2014,Fox2015,Sarkar2015,Ko2019}. X-ray observations show that the edges of the Fermi bubbles coincide well with the biconical X-ray structure discovered in the ROSAT 1.5 keV map, indicating that they could have the same origin \citep[][]{BlandHawthorn2003,Keshet2017,Keshet2018,BlandHawthorn2019}. X-ray observations thus put a strong constraint on the origin of the Fermi bubbles, and recently \citet[][hereafter Z20]{Zhang2020} show that the gas density, temperature, and X-ray surface brightness of the biconical X-ray structure could be naturally reproduced by the forward shock driven in the MW CGM by an AGN jet event occurring about 5 Myr ago. \citetalias{Zhang2020} further shows that the narrow biconically-shaped base of the Fermi bubbles could not be reproduced by forward shocks driven by spherical AGN or stellar feedback outflows from the GC.

X-ray observations of the Fermi bubbles are also a useful diagnostic probe for the gas density distribution in the inner Galaxy, suggesting that the inner Galaxy contains a large gaseous shell heated and swept up by a shock front (\citealt{BlandHawthorn2003}; \citetalias{Zhang2020}). In this paper, we combine hydrodynamic simulations and X-ray observations of the Fermi bubbles and investigate the halo gas distribution in the inner Galaxy before the Fermi bubble event. This gas distribution supplies an initial condition of the environment for the formation of the Fermi bubbles, and may smoothly extend to the large-scale MW CGM. Alternatively, it may also have been substantially disturbed by previous energetic events at the GC, and may thus provide important clues on feedback processes prior to the Fermi bubble event. We present a series of simulations with a variety of initial gas distribution models in the inner Galaxy and investigate the evolution of the Fermi bubbles in the AGN jet scenario where the bubble edge corresponds to the jet-triggered forward shock. We use the X-ray surface brightness and the morphology of the Fermi bubbles as a diagnostic for the initial halo gas distribution. The Fermi bubbles thus become a unique laboratory to study the gas distribution in the inner Galaxy.

The rest of this paper is organized as follows. In Section 2, we list the halo density models explored in this work. In Section 3, we describe the numerical setup of our simulations. The results are presented in Section 4. In Section 5, we discuss the structure of the MW CGM inferred from our results and the impact of gas metallicity and temperature in the Fermi bubbles on our results. We summarize our results in in Section 6.

\section{Models of the Gaseous Halo}\label{sec-models}

A variety of the hot gas distribution models for the MW CGM have been proposed in the literature. Here we investigate several representative ones, including five spherical models and two disk-like models. Throughout this paper, the hot gas density $n_{\rm e}$ specifically refers to the thermal electron number density in the hot gaseous halo. A brief description of the spherical density models are listed as follows:

\begin{enumerate}
\item The isothermal hydrostatic model. This model assumes that the hot gaseous halo is spatially isothermal with the thermal pressure gradient balancing gravity, and has been widely used in numerical simulations \citep[e.g.][]{Guo2012,Yang2012,Mou2014,Sarkar2015}. Here we adopt the same isothermal model previously used in \citetalias{Zhang2020} to describe the halo gas distribution (hereafter the Z20 model). The hot gas is assumed to be in hydrostatic equilibrium with temperature $T = 2.32\times 10^{6}$ K ($0.2$ keV) in the Galactic potential, which is determined by an updated MW mass model proposed in \citet[][see \citetalias{Zhang2020} for details]{McMillan2017}. The density profile is then derived from hydrostatic equilibrium with the normalization thermal electron number density at the GC set to be $n_{0}=3.0\times 10^{-2}$ cm $^{-3}$. As shown in \citetalias{Zhang2020}, the derived gas density profile agrees quite well with the spherical best-fit $\beta$ model presented in \citet{Miller2015} at $r\gtrsim 2$ kpc.

 \item The $\beta$ model. The isothermal gas density distribution under hydrostatic equilibrium in the Navarro-Frenk-White (NFW; \citealt{Navarro1996}) dark matter potential can be roughly approximated by the spherical $\beta$ model described below \citep{Cavaliere1976,Makino1998}:
    \begin{equation}\label{beta}
    n_{\rm e}(r)=n_{0}(1+(r/r_{c})^2)^{-3\beta/2},
    \end{equation}
where $n_{0}$ is the normalization factor (as in other halo gas models hereafter), $r$ is the distance to the GC, $r_{c}$ is the core radius, and $\beta$ determines the slope. For the MW, a flattened axisymmetric density profile $n(R,z)$ may be expected \citep{Stewart2013}:
    \begin{equation}\label{beta3}
    n_{\rm e}(R,Z)=n_{0}[1+(R/R_{c})^2+(z/z_{c})^2]^{-3\beta/2},
    \end{equation}
where $(R,z)$ are the coordinates in the cylindrical coordinate system centered at the GC, and the $z$ axis follows the Galaxy rotation axis. The \citetalias{Zhang2020} model in the inner Galaxy can be roughly approximated by a nearly-spherical $\beta$ model (Equation \ref{beta3}) with $R_{c}=0.58$ kpc, $z_{c}=0.45$ kpc, $n_{0}=0.03$ cm$^{-3}$ and $\beta=0.5$. In this paper, we investigate this axisymmetric $\beta$ model and perform a parameter study over $\beta$ (see Table 1). We note that this axisymmetric $\beta$ model with $\beta=0.5$ agrees quite well with the spherical $\beta$ model in \citet{Miller2015} at $r\gtrsim 2$ kpc.

 \item The MB model \citep{Maller2004}. This model assumes that the halo gas is adiabatic and under hydrostatic equilibrium in the NFW dark matter potential of the MW. The gas density profile $n_{\rm e}(x)$ in this model is (\citealt{Maller2004}; \citealt{Fang2013}):
  \begin{equation}\label{eq-mb}
  n_{\rm e}(x)=n_{0}[1+\frac{3.7}{x}\mathrm{ln}(1+x)-\frac{3.7}{C_{\mathrm{v}}}\mathrm{ln}(1+C_{\mathrm{v}})]^{3/2},
  \end{equation}
where $x\equiv r/r_{\rm s,MB}$. $r_{\rm s,MB}$ and $C_{\mathrm{v}}$ are the scale radius and the concentration parameter of the MW dark matter distribution, respectively. In our simulation, we adopt the parameters presented in \citet{Fang2013}, where $C_{\mathrm{v}}=12$ and $r_{\rm s,MB}=21.7$ kpc. The temperature profile in the inner MW halo in this model is quite flat, and for simplicity, we set $T=0.2$ keV, which is consistent with recent observations \citep{Kataoka2018}. We choose $n_{0}=1.8\times 10^{-5}$ cm $^{-3}$ so that the resulted gas density profile (Equation \ref{eq-mb}) roughly matches the MB model presented in \citet{Fang2013}.

 \item The NFW model. This model simply assumes that the hot gas density distribution traces the dark matter density distribution:
   \begin{equation}\label{NFW}
    n_{\rm e}(x)=\frac{n_{0}}{x(1+x)^{2}},
   \end{equation}
where $x\equiv r/r_{\rm s,NFW}$. $r_{\rm s,NFW}$ is the scale radius of the dark matter NFW profile, and we adopt $r_{\rm s,NFW}=32.5$ kpc as in \citet{Fang2020}. The normalization parameter $n_{0}$ is set to be $n_{0}=1.1\times 10^{-3}$ cm$^{-3}$, which is obtained by fitting the profile with two recent electron number density estimates from the ram-pressure stripping calculations of the MW satellite galaxies: $n_{\rm e} =(6.8-18.8)\times 10^{-5}$ cm$^{-3}$ at $r =70 \pm 20$ kpc \citep{Gatto2013} and $n_{\rm e} =(3.4-8.0) \times 10^{-5}$ cm$^{-3}$ at $r = 48.2 \pm 2.5$ kpc \citep{Salem2015}. Here we have converted the estimated total number densities in these two references to $n_{\rm e}$ (see \citealt{Guo2020}).

 \item The \citet{Guo2020} model (hereafter denoted as the G20 model). Recently, \citet{Guo2020} proposed a physically-motivated model where the density distribution of the hot halo gas follows a generic analytic form:
     \begin{equation}\label{eq-Guo}
     n_{\rm e}(r)=n_{0}\frac{r_{2}^{3}}{(r+r_{1})^{\alpha}(r+r_{2})^{3-\alpha}},
    \end{equation}
 where $r_{1}$ represents the size of an inner density core and $r_{2}$ represents the impact of Galactic feedback processes on the halo gas distribution. If $\alpha=1$, $r_{1}=0$ and $r_{2}=R_{s}$, Equation \ref{eq-Guo} reduces to the NFW profile. If $\alpha=1.5$ and $r_{1}=0$, this model reduces to the $\beta$ model with $\beta=0.5$ at $r_{c}\ll r \ll r_{2}$. Here we investigate a cored G20 model \citep{Fang2020} with $\alpha=1$, $r_{1}=23.4$ kpc, and $r_{2}=100$ kpc, where the gas density distribution in the inner Galaxy is relatively flat. $n_{0}=1.93\times 10^{-4}$ cm$^{-3}$ is determined in the same way as in the NFW model above.

\end{enumerate}

 \begin{figure}[h!]
 \centering
 \gridline{
  \includegraphics[width=8cm]{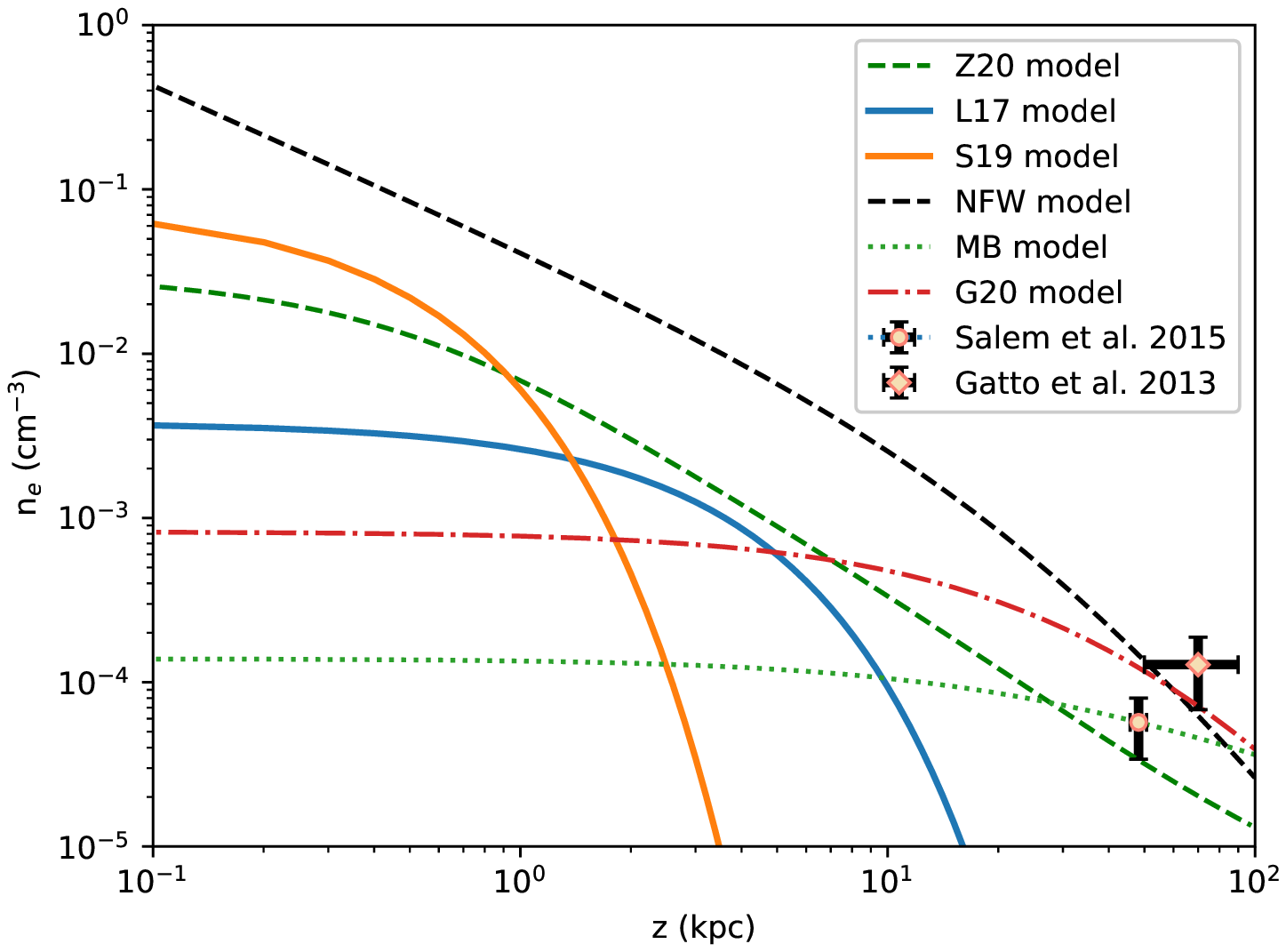}
  }
  \gridline{
  \includegraphics[width=8cm]{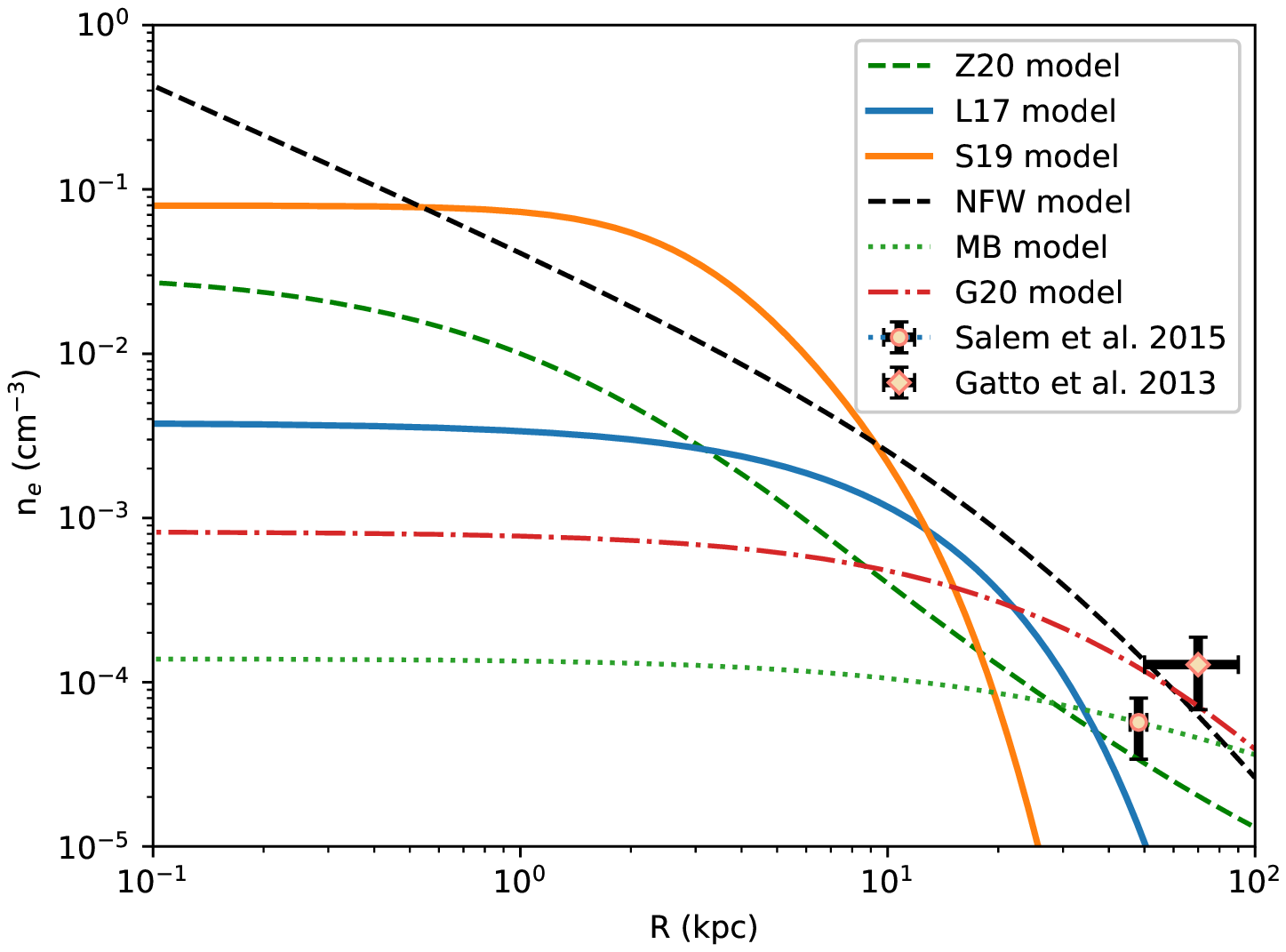}
  }
\caption{The initial density distributions of the hot halo gas along the $z$ axis (top) and the $R$ axis (bottom) in a variety of models described in Section \ref{sec-models}, including the NFW model, the Z20 model \citep{Zhang2020}, the MB model, the G20 model \citep{Guo2020}, the S19 disk-like model \citep{Sofue2019}, and the L17 disk-like model \citep{Li2017}. The upper and lower data points correspond to the recent density estimates by \citet{Gatto2013} and \citet{Salem2015}, respectively (also see \citealt{Guo2020}).
}\label{fig-den1d}
 \end{figure}

Disk-like density models have also been proposed and investigated for the MW CGM. \citet{Yao2007} proposed a nonisothermal disk-like model with the gas density and temperature decreasing exponentially along the $z$ direction. \citet{Li2017} revised this model by further assuming that the gas density also decreases exponentially along the $R$ direction. Disk-like models have also been recently adopted to study the X-ray emission measure distribution along different sight lines \citep{Nakashima2018,Kaaret2020}. Recently, \citet{Sofue2019} adopted a new disk-like model to study the formation of the North Polar Spurs (NPS) and South Polar Spurs (SPS) in the MW halo. In the current work, we choose the models in \citet{Li2017} and \citet{Sofue2019} as two representative disk-like models, which are described below:

\begin{enumerate}
 \item The \citet{Li2017} disk-like model (hereafter denoted as the L17 model). In this model, the gas density $n(R,z)$ decrease along both the $R$ and $z$ directions:
 \begin{equation}\label{eq_LB_disk}
 n_{\rm e}(R,z)=n_{0}e^{-R/R_{h}}e^{-z/z_{h}}
 \end{equation}
 where $R_{h}$ and $z_{h}$ are the scale length in the Galactic plane and the scale height along the $z$ axis, respectively. We adopt the model parameters from \citet{Nakashima2018}: $n_{0}=3.8\times 10^{-3}$ cm$^{-3}$, $R_{h}=7.0$ kpc and $z_{h}=2.7$ kpc.

 \item The \cite{Sofue2019} disk-like model (hereafter the S19 model). This model can be used to describe a rotating gaseous disk in the Galactic potential \citep[see][]{Sormani2018}. The gas density profile in this model is
 \begin{equation}\label{eq_Sofue_disk}
 n_{\rm e}(R,z)=n_{0,S}e^{-|z|/z_{X}(R)}A(R),
 \end{equation}
 where
 \begin{equation}\label{eq_Sofue_disk2}
  z_{X}(R)=z_{X}(R_{0})\frac{S(R_{0}/a_{d})}{S(R/a_{d})}
 \end{equation}
 and
 \begin{equation}\label{eq_Sofue_disk3}
  A(R)=\frac{S(R/R_{x})}{S(R_{0}/R_{x})}\frac{z_{X}(R_{0})}{z_{X}(R)}.
 \end{equation}
 Here $S(x)$ is a semi-exponential function,
 \begin{equation}\label{sx}
  S(x)=fe^{-x}+(1-f)e^{-x^{2}}
 \end{equation}
 where $f=x^{2}/(\alpha^{2}+x^{2})$ with $\alpha=0.5$. Following \citet{Sofue2019}, we adopt $n_{0,S}=3.7 \times 10^{-3}$ cm$^{-3}$, $R_{0}=8.5$ kpc, $z_{X}(R_{0})=1.8$ kpc, $R_{x}=a_{d}=5.73$ kpc. A uniform background density $10^{-5}$ cm$^{-3}$ is added in this model as in \cite{Sofue2019}.
\end{enumerate}

The CGM density models described above show large differences in the inner Galaxy, as clearly shown in Figure \ref{fig-den1d}. The gas density in the central kiloparsec varies about three orders of magnitude, from $10^{-1}$ cm$^{-3}$ in the NFW model to $10^{-4}$ cm$^{-3}$ in the MB model. The density slope in the inner Galaxy also varies significantly in different models. While the density profile in the inner Galaxy is relatively flat in the MB and G20 models, it decreases dramatically along the $z$ axis in the S19 model. We note that in the disk-like L17 and S19 models, the gas densities at large radii ($r\sim 50-90$ kpc) are significantly lower than the recent density estimates from the ram-pressure stripping models by \citet{Gatto2013} and \citet{Salem2015}, as shown in Figure \ref{fig-den1d}. These differences show large uncertainties in our understanding of the unperturbed halo gas distribution in the inner Galaxy.

The temperature of the hot MW halo gas has also been measured by X-ray observations. \citet{Yoshino2009} analyzed the OVIII and OVII emission lines observed by Suzaku and found that the hot gas temperature shows a narrow range around $\sim 0.2$ keV. \citet{Henley2013} investigated X-ray emissions along 110 XMM-Newton sight lines and found that the median temperature is $0.19 $ keV with an interquartile range of about $0.05$ keV. Recent X-ray observations by \citet{Nakashima2018} and \citet{Kaaret2020} lead to slightly higher halo gas temperatures, possibly due to their different treatments on the emission of the local hot bubble. These measured temperatures are line-of-sight averaged gas temperatures and are quite uniform along different sight lines. In this paper, we focus on the halo gas distribution in the inner Galaxy, and for simplicity, we choose a spatially constant temperature $T = 2.32\times 10^{6}$ K ($0.2$ keV) for the unperturbed halo gas before the Fermi bubble event. We note that, as demonstrated by \citet{Guo2020}, a radially-variable temperature profile could also lead to relatively uniform temperature measurements along different sight lines, due to the fact that the Earth is located quite close to the center of the large MW halo.

\section{Simulation setup} \label{sec3}

\begin{deluxetable*}{ccccccccccc}\label{tab-para}
\tablecaption{List of our Simulations with Model Parameters and Some Key Results.}
\tablewidth{0pt}
\tablehead{
\colhead{Run}      &\colhead{CGM model}   &$\rho_{j}$        &$e_{j}$  &\colhead{$v_{j}$}    &\colhead{$e_{\rm jcr}$}   &\colhead{$P_{\rm j}$} &\colhead{$t_{\rm FB}$}  &\colhead{$E_{\rm j}$} &\colhead{$f$} \\
\colhead{ID}           &  &g cm$^{-3}$   &erg cm$^{-3}$ &\colhead{$10^{9}$cm $s^{-1}$}  & \colhead{erg cm$^{-3}$}        & \colhead{erg~s$^{-1}$}  &\colhead{Myr}       &\colhead{erg} &}
\startdata
A     &Z20\citep{Zhang2020} &$1.23\times10^{-27}$ & $1.46\times10^{-11}$  &$2.5$  &$2.7\times10^{-10}$  &$3.42\times10^{41}$ &5 &$1.07\times10^{55}$     &1.31\\
B     &MB\citep{Fang2013}  &$1.06\times10^{-29}$ &$0.19\times10^{-12}$   &$5.0$  &$8.0\times10^{-12}$  &  $2.34\times10^{40}$&5 &$7.38\times10^{53}$ &15.12\\
C    &G20\citep{Guo2020} &$6.26\times10^{-29}$ &$7.40\times10^{-13}$  &$4.0$  &$8.0\times10^{-11}$  &$2.40\times10^{41}$ &5 &$7.56\times10^{54}$           &3.25\\
D    &NFW\citep{navarro97}  & $9.19\times10^{-27}$& $1.09\times10^{-10}$  &$2.3$  &$2.0\times10^{-9}$      &$2.01\times10^{42}$&5 & $6.34\times10^{55}$  &0.25\\
E    &L17\citep{Li2017}  & $2.58\times10^{-28}$& $3.06\times10^{-12}$  &$0.9$  &$2.0\times10^{-10}$      &$3.01\times10^{40}$&5 & $9.47\times10^{54}$  &1.60\\
F    &S19\citep{Sofue2019} & $2.11\times10^{-27}$& $2.49\times10^{-11}$  &$0.8$  &$7.0\times10^{-10}$      &$3.71\times10^{40}$&5 & $1.17\times10^{54}$      &1.05\\
G    &$\beta$ model, $\beta=0.3$  & $1.87\times10^{-27}$& $2.21\times10^{-11}$  &$3.5$  &$8.0\times10^{-10}$   &$1.42\times10^{42}$&5 & $4.17\times10^{55}$       &0.42\\
H    &$\beta$ model, $\beta=0.4$  & $1.73\times10^{-27}$& $2.06\times10^{-11}$  &$2.6$  &$4.0\times10^{-10}$ &$5.41\times10^{41}$&5 & $1.70\times10^{55}$          &0.76\\
I    &$\beta$ model, $\beta=0.6$  & $1.50\times10^{-27}$& $1.77\times10^{-11}$  &$1.9$  &$3.0\times10^{-10}$       &$1.90\times10^{42}$&5 & $5.98\times10^{54}$   &1.99\\
J    &$\beta$ model, $\beta=0.7$  & $1.39\times10^{-27}$& $1.65\times10^{-11}$  &$1.7$  &$2.5\times10^{-10}$     &$1.28\times10^{41}$&5 & $4.03\times10^{54}$ &3.00\\
\enddata
\tablecomments{In each simulation, the jet is initialized with the following parameters: gas density $\rho_{j}$, thermal energy density $e_{j}$, CR energy density $e_{\rm jcr}$, velocity $v_{j}$. $P_{\rm j}$ and $E_{\rm j}$ refer to the power and the total injected energy of one jet, respectively. The jet lasted for 1 Myr in all runs. $t_{\rm FB}$ is the current age of the Fermi bubbles in each simulation. $f$ is the rescaling factor used to fit the ROSAT 1.5 keV X-ray surface brightness distribution of the Fermi bubbles (see Section \ref{sec-x}). A brief description of our simulations are given in Section \ref{sec3}.}
\end{deluxetable*}

The numerical setup of our simulations is essentially the same as in our previous work \citetalias{Zhang2020}, except that in the current paper, we investigate a variety of different halo gas density models described in Section \ref{sec-models}. We adopt the same 2-dimensional (2D) Eulerian grid-based hydrodynamic code presented in \citetalias{Zhang2020}, which was also used in \citet{Guo2012} and \citet{Guo2012a}. Assuming axisymmetry around the MW rotation axis, we solve hydrodynamic equations in cylindrical coordinates $(R, z)$, where the $z$ axis is along the MW rotation axis. AGN jets contain both thermal gas and cosmic rays (CRs), and CRs are treated as a second fluid in our simulations with the adiabatic index of $\gamma_{c}=4/3$. Along each coordinate axis, the simulation domain consists of 1800 uniformly-spaced zones from 0 to 15 kpc and 100 logarithmically spaced zones from 15 to 70 kpc. We adopt outflow boundary conditions at the outer boundaries and reflective boundary conditions at the inner boundaries.

To set up the MW gravitational potential, we adopt the best-fitting MW mass model proposed in \citet[][see \citetalias{Zhang2020} for more details]{McMillan2017}. In this model, six mass components are considered, including the dark matter halo, the Galactic bulge, the thin and thick stellar disks, and the HI and molecular gas disks. We perform a series of hydrodynamic simulations with various initial halo gas density distributions as described in Section \ref{sec-models}, and the model parameters of these simulations are listed in Table \ref{tab-para}. The initial halo gas temperature is set to be $T = 2.32\times 10^{6}$ K in all our simulations. The initial halo gas is in hydrostatic equilibrium in runs A and H, where the Z20 density model is adopted as in \citetalias{Zhang2020}. In all other runs, the initial halo gas is not in hydrostatic equilibrium, which does not have substantial impacts on our results due to the relatively short jet duration and the simulation time. In reality, the halo gas in the inner Galaxy may indeed not be in hydrostatic equilibrium due to AGN and stellar feedback processes from the GC \citep{Oppenheimer2018}.

We adopt the same jet initialization method from \citetalias{Zhang2020}. The jet is launched at $t=0$ along the $z$ axis from a cylinder at the GC with radius $R_{j}=33.3$ pc and length $z_{j}=350$ pc. At the base, the jet contains both thermal gas and CRs with energy densities $e_{j}$ and $e_{jcr}$, respectively. $e_{j}$ and $e_{jcr}$ are largely degenerate with respect to jet evolution, and in our simulations, we set $e_{j}$ equal to the ambient gas energy density while leaving $e_{jcr}$ as a free parameter to fit the observed morphology of the Fermi bubbles in the gamma-ray band. At the base, the mass density and velocity of the jet are $\rho_{j}$ and $v_{j}$, respectively. The jet is active for a duration of $t_{\rm j} = 1.0$ Myr.

For each halo gas model, we perform a large number of simulations, scanning through the jet parameter space $(\rho_{j}, v_{j}, e_{jcr})$. As in \citetalias{Zhang2020}, the best-fit run for each CGM model is selected according to two observational constraints: the bilobular morphology of the Fermi bubbles and the current temperature $\sim 0.4$ keV \citep{Miller2016} of the shock-compressed gas in the Fermi bubbles. Each simulation is stopped at $t=t_{\rm FB}$, when the size of the simulated Fermi bubble reaches the observed bubble size. The jet parameters and some key results of our best-fit runs for all our CGM models are listed in Table \ref{tab-para}.

Run A is the fiducial run using the cuspy \citetalias{Zhang2020} model for the initial CGM density distribution. The main results of run A have been presented in \citetalias{Zhang2020}, and here we focus on the X-ray emission properties of the Fermi bubbles. Adopting the MB and G20 models respectively, runs B and C start with a cored initial halo gas density distribution. Run D adopts the centrally cuspy NFW profile as the initial CGM density distribution. We use runs E and F to investigate the disk-like L17 and S19 models of the CGM density distribution, respectively. Runs G, H, I, and J are presented to investigate the slope of the halo gas density profile in the inner Galaxy, using the $\beta$ model with $\beta = 0.3$, $0.4$, $0.6$, and $0.7$, respectively. As shown in \citetalias{Zhang2020}, the \citetalias{Zhang2020} model in run A can be approximated by a $\beta$ model with $\beta = 0.5$.

\section{Results}

For each simulation, we calculate the distributions of the X-ray surface brightness and the O VIII/O VII emission line ratio of the Fermi bubbles. By comparing the simulation results directly with the ROSAT X-ray surface brightness distribution from \citet{Snowden1997} and the O VIII/O VII emission line ratios from archival XMM-Newton and Suzaku data \citep{Miller2016}, we then constrain the density distribution of the unperturbed MW CGM before the Fermi bubble event.

\subsection{The Fiducial Run}

\begin{figure}[h!]
 \centering
  \gridline{
  \includegraphics[width=8cm]{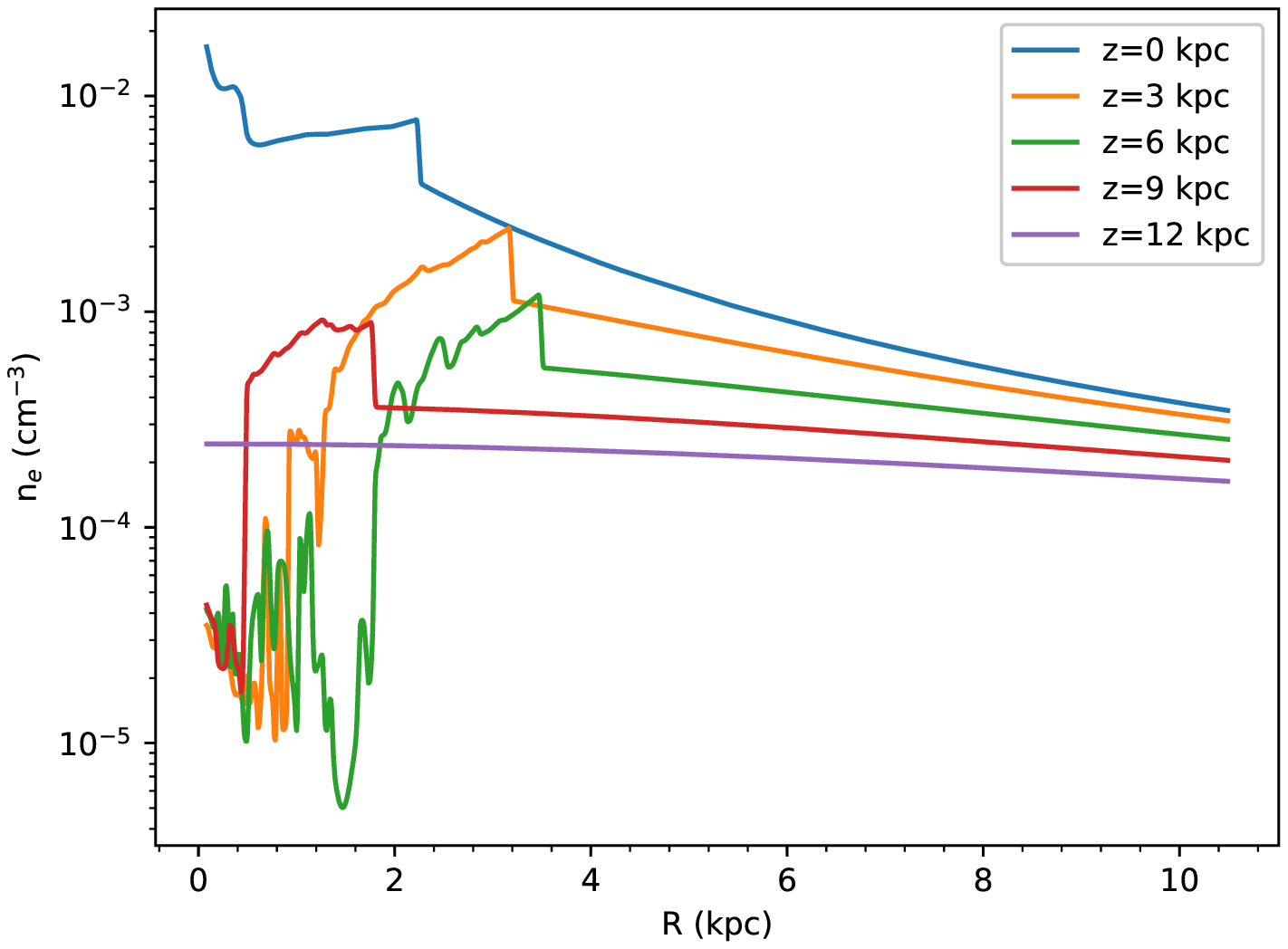}}
 \gridline{
  \includegraphics[width=8cm]{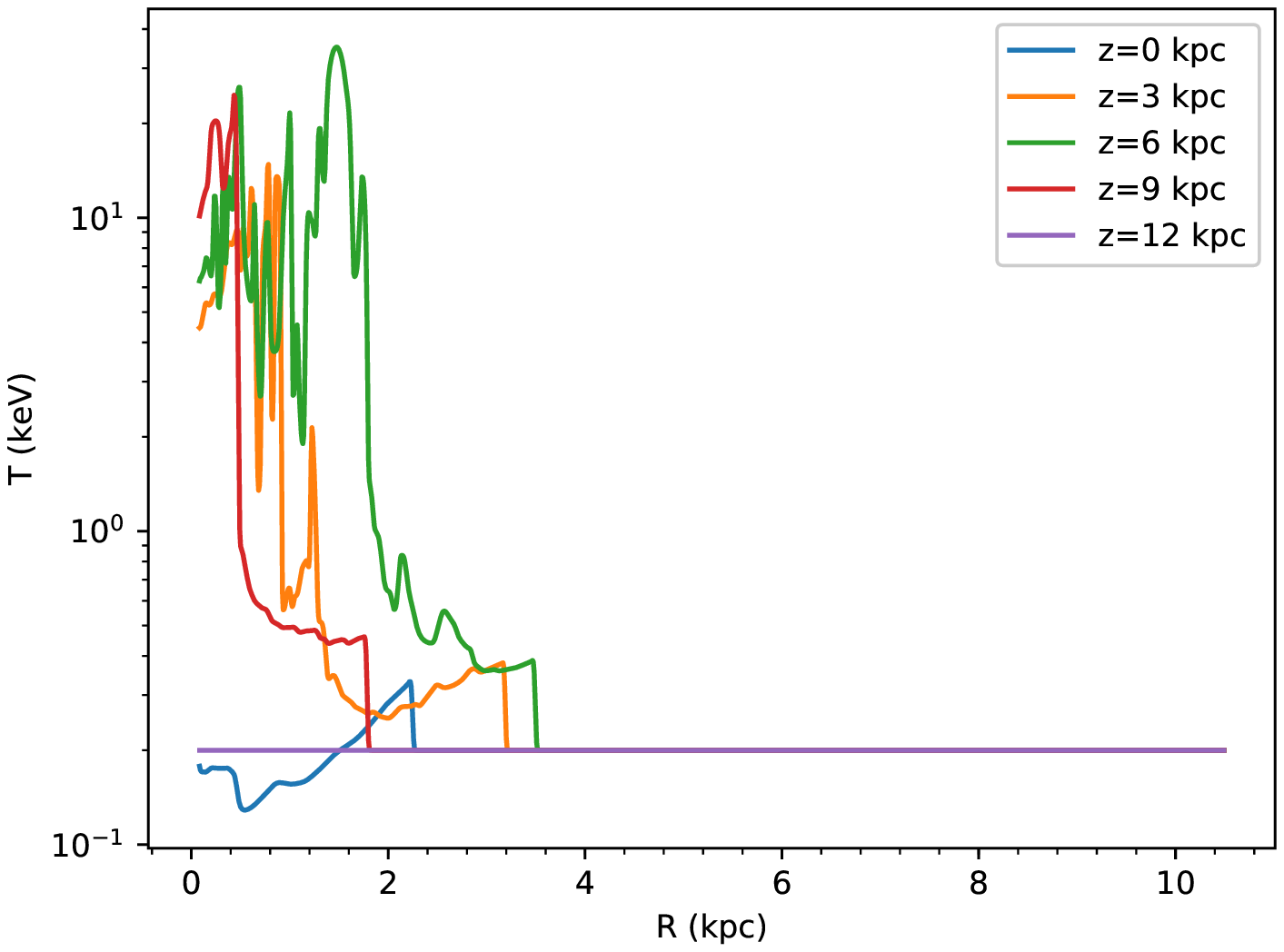}}
\caption{ Variations of thermal gas density (top) and temperature (bottom) along the $R$ direction in run A at $t = 5$ Myr at five fixed values of $z = 0$, $3$, $6$, $9$, and $12$ kpc. }\label{fig-denT1d}
 \end{figure}

In this subsection, we show the results of the fiducial run---run A, which has been investigated in details in \citetalias{Zhang2020} on the evolution of the Fermi bubbles. Here we focus on X-ray emissions of the bubbles at the current time, i.e., $t=t_{FB}$, which is $5$ Myr in run A. We first show in Figure \ref{fig-denT1d} the distributions of electron number density (top) and temperature (bottom) of thermal gas at $t=5$ Myr. The sharp jump along each line at $R= 2-4$ kpc is the location of the forward shock, corresponding to the edge of the bubbles. In the shock downstream is a shell of shock-compressed halo gas. The jet ejecta is located in the innermost part of the bubbles with very high temperatures up to $T\sim 10$ keV. The gas density in the ejecta is on the order of $10^{-5}$ cm$^{-3}$, about two orders of magnitude lower than the gas density in the compressed shell near the bubble edge. For optically thin plasma in collisional ionization equilibrium, the X-ray emissivity per unit volume is proportional to the square of the gas density. The X-ray emissivity in the ejecta is thus about four orders of magnitude lower than that in the outer shock-compressed shell. Therefore, the X-ray emission of the Fermi bubbles is dominated by the outer shock-compressed shell, while the contribution from the jet ejecta is negligible.

\subsubsection{The X-ray Surface Brightness Distribution}\label{sec-x}

\begin{figure}[h!]
 \centering
  \includegraphics[width=8cm]{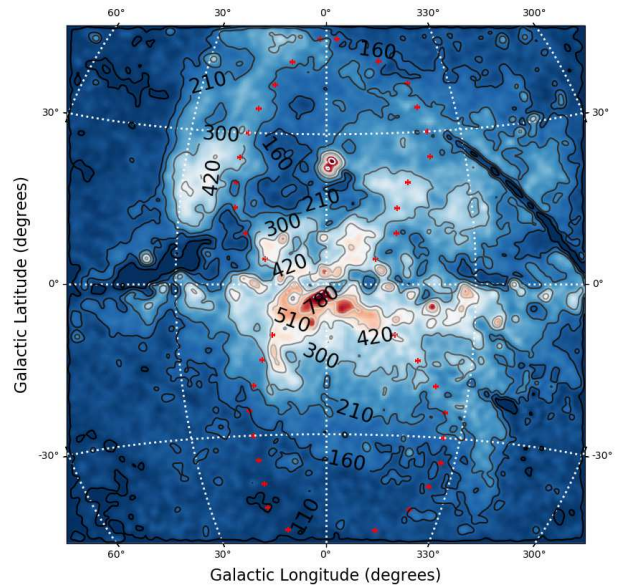}
\caption{Contour map of the ROSAT 1.5 keV (0.73--2.04 keV) X-ray surface brightness distribution smoothed by a Gaussian filter using the open-source python package ASTROPY with the standard deviation stddev $= 4$. The unit is $10^{-6}$ counts s$^{-1}$ arcmin$^{-2}$. The red dots represent the edges of the observed Fermi bubbles in gamma rays.}\label{fig-R6R7}
 \end{figure}

For the observed X-ray surface brightness distribution, we adopt the ROSAT all-sky survey in the 1.5 keV band (0.73--2.04 keV; see \citealt{Snowden1997} for the survey details). Figure \ref{fig-R6R7} shows the ROSAT 1.5 keV X-ray surface brightness map of the inner Galaxy region, clearly indicating a biconical X-ray structure at the GC as first described in \citet{BlandHawthorn2003}. The X-ray structure is centered at about the GC, symmetric about the Galactic plane, and bright mainly in the low-latitude region of $|l|<20^{\circ}$ and $|b|<20^{\circ}$. Figure \ref{fig-R6R7} clearly shows that the outline of this biconical X-ray structure spatially coincides with the edges of the Fermi bubbles very well at low latitudes, which implies that the two structures share the same origin and may be naturally explained by AGN jet-driven forward shocks as demonstrated by \citetalias{Zhang2020}. The X-ray surface brightness decreases roughly from $6 \times 10^{-4}$ near the Galactic plane to $2 \times 10^{-4}$ counts s$^{-1}$ arcmin$^{-2}$ at $|b|\sim20^{\circ}$ in both the northern and southern Galactic hemispheres, and the X-ray structure becomes invisible at $|b|>30^{\circ}$, indicating lower gas densities at higher values of $z$.

In run A, we use the simulated gas density and temperature distributions at $t= 5$ Myr to make the synthetic $0.7-2.0$ keV X-ray surface brightness map $I(l,b)$ in Galactic coordinates $(l,b)$,
\begin{equation}\label{brightness}
I(l,b)=\frac{1}{4\pi}\int_{\rm los} n_{\rm H}n_{\rm e}\epsilon(T,Z)dl ~~ \textrm{erg s}^{-1} \textrm{cm}^{-2} \textrm{sr}^{-1}.
\end{equation}
where $n_{\rm H}$ and $n_{\rm e}$ represent the hydrogen and electron number densities in the hot plasma, respectively. $\epsilon(T,Z)$ is the $0.7-2.0$ keV X-ray emissivity adopted from the APEC plasma model \citep{Smith2001,Foster2012} with AtomDB (version 3.0.9). Both line and continuum emissions are included. $T$ is the gas temperature, and a uniform metallicity of $Z=0.4Z_{\odot}$ is adopted. We assume that the distance from the Sun to the GC is 8.5 kpc and the integral is done to a distance of 50 kpc along any line of sight. In order to directly compare with ROSAT observations (Fig. \ref{fig-R6R7}), the unabsorbed energy flux $I(l,b)$ in the $0.7-2.0$ keV band is further converted to the photon flux in the ROSAT $0.73-2.04$ keV (R6-R7) band with the PIMMS utility \footnote{\url{https://heasarc.gsfc.nasa.gov/cgi-bin/Tools/w3pimms/w3pimms.pl}} (v4.10) assuming the APEC plasma model with $T=0.385$ keV and $Z = 0.4 Z_{\odot}$. We also take into account the absorption by neutral hydrogens by assuming that the HI column density along any sight line from the Sun to the Fermi bubbles is half of the corresponding value from the Effelsberg-Bonn HI survey \citep{Winkel2016}.

\begin{figure}[h!]
 \centering
  \includegraphics[width=8cm]{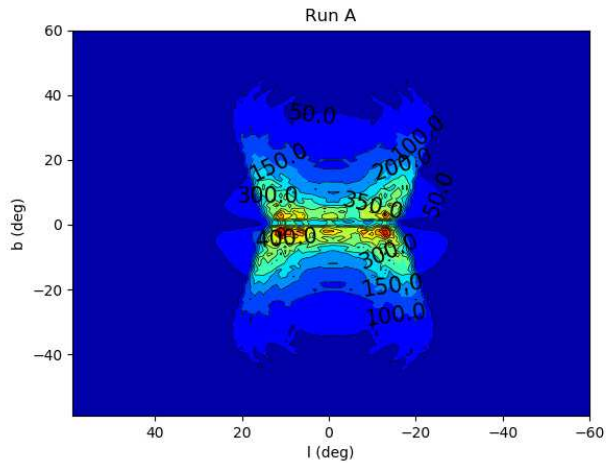}
\caption{Contour map of the synthetic $0.73-2.04$ keV X-ray surface brightness distribution in units of $10^{-6}$ counts s$^{-1}$ arcmin$^{-2}$ in run A at $t= 5$ Myr in Galactic coordinates. The ROSAT instrument response and the absorption of neutral hydrogens have already been taken into account using the PIMMS utility.} \label{fig-x-runA2d}
 \end{figure}

The derived synthetic X-ray surface brightness (photon flux) map is shown in Figure \ref{fig-x-runA2d}. The biconical X-ray structure near the GC in the ROSAT map shown in Figure \ref{fig-R6R7} is well reproduced. The limb-brightening feature also indicates that the X-ray emission is dominated by a shell within the edge of the Fermi bubbles. Note that until now we have not tuned any model parameters with the ROSAT data, and remarkably the observed X-ray surface brightness of the Fermi bubbles is almost naturally reproduced. The X-ray surface brightness at $|b| \gtrsim 30^{\circ}$ is lower than the background emission ($ 7.0\times10^{-5}$ counts s$^{-1}$ arcmin$^{-2}$), which explains why the Fermi bubbles are invisible at high latitudes in ROSAT X-ray observations.

\begin{figure}[h!]
 \centering
  \includegraphics[width=8cm]{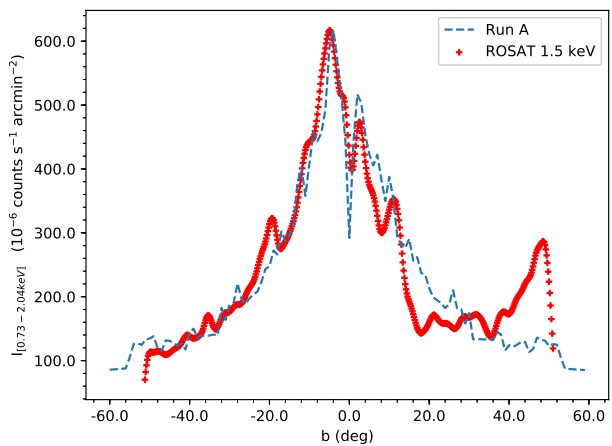}
\caption{Simulated $0.73-2.04$ keV X-ray surface brightness profile as a function of Galactic latitude at a fixed Galactic longitude $l=5^{\circ}$ in run A at $t= 5$ Myr, rescaled by a factor of $f^{2}=1.31^{2}$ to fit the ROSAT 1.5 keV X-ray data (red pluses). }\label{fig-x-RunA1d}
 \end{figure}

A quantitative comparison between the observed and synthetic X-ray surface brightness distributions of the Fermi bubbles could be used to constrain the unperturbed CGM density profile. Figure \ref{fig-x-RunA1d} shows the $0.73-2.04$ keV surface brightness profile as a function of Galactic latitude at a fixed Galactic longitude $l=5^{\circ}$ in run A at $t= 5$ Myr. We choose $l=5^{\circ}$ rather than $l=0^{\circ}$ to avoid the pollutions by Sgr A* and Sco X-1 along $l \sim 0^{\circ}$. { A uniform photon flux of $7.0 \times 10^{-5}$ counts s$^{-1}$ arcmin$^{-2}$ (which is the minimum flux along $l=5^{\circ}$ in the ROSAT data)} is added to mimic the diffuse background and foreground X-ray emissions. We rescale the simulated X-ray surface brightness profile in Figure \ref{fig-x-RunA1d} by a factor of $f^{2}$ so that the maximum synthetic surface brightness $I_{syn}$ along the longitude $l=5^{\circ}$ equals to the maximum surface brightness in the ROSAT data $I_{obs}$ along the same longitude:
\begin{equation}\label{cali}
f^{2}=\frac{\textrm{max}(I_{obs})}{\textrm{max}(I_{syn})}.
\end{equation}
A rescaling factor of $f^{2}$ in X-ray surface brightness corresponds to a rescaling factor of $f$ in the initial MW CGM density model. Since the hydrodynamic equations governing the jet evolution is scale free (see \citetalias{Zhang2020}), the morphological evolution of the simulated Fermi bubbles is unchanged if the jet parameters $\rho_{j}$, $e_{j}$ and $e_{jcr}$ are changed by the same factor of $f$ (keeping the other jet parameters fixed). We perform the same analysis for all the runs in Table \ref{tab-para}, which allows us to use the ROSAT observations to constrain the CGM density models in Section \ref{sec-models} to be $f n$. The derived values of $f$ for all our runs are shown in the rightmost column of Table \ref{tab-para}. It is interesting to note that, after calibrating with the ROSAT data, the rescaled jet energy $f E_{\rm j}$ is about $10^{55}$ erg for most runs.

For run A,  we have $f =1.31$, indicating that the thermal electron number density at the GC in the Z20 model is $f n_{0}=0.04$ cm $^{-3}$ and the jet energy is $f E_{\rm j}=1.4\times 10^{55}$ erg. Figure \ref{fig-x-RunA1d} clearly shows that the simulated X-ray surface brightness profile in run A with the Z20 density model provide a reasonably good fit to the ROSAT data. The decrease of the X-ray surface brightness with Galactic latitude indicates that the gas density in the shock-compressed shell within the Fermi bubbles decreases with $z$, which results from the same trend in the initial CGM density distribution. The synthetic surface brightness profile matches the ROSAT data better in the southern hemisphere ($b<0^{\circ}$) than in the northern hemisphere ($b>0^{\circ}$), which could be attributed to additional X-ray emissions and absorptions in the northern hemisphere. At $b\sim 0^{\circ}$, the drop in the ROSAT data is reproduced, suggesting that the drop is caused by the absorption of the HI gas located between the Sun and the Fermi bubbles. The drop at $b\sim10^{\circ}$ in Figure \ref{fig-x-RunA1d} corresponds to the shadow of the Ophiuchus dark clouds, and the bump at $b\sim50^{\circ}$ is due to the X-ray emission of the NPS. The drop at b=$12^{\circ}-28^{\circ}$ in Figure \ref{fig-x-RunA1d} corresponds to the low X-ray surface brightness region at $0^{\circ}<l<15^{\circ}$ and $15^{\circ}<b<40^{\circ}$ in Figure \ref{fig-R6R7}, and may be caused by the absorption of some foreground dusty clouds (probably the Serpens-Aquila Rift).

\subsubsection{Emission Line Ratio of O VIII to O VII}

\begin{figure}[h!]
 \centering
  \includegraphics[width=8cm]{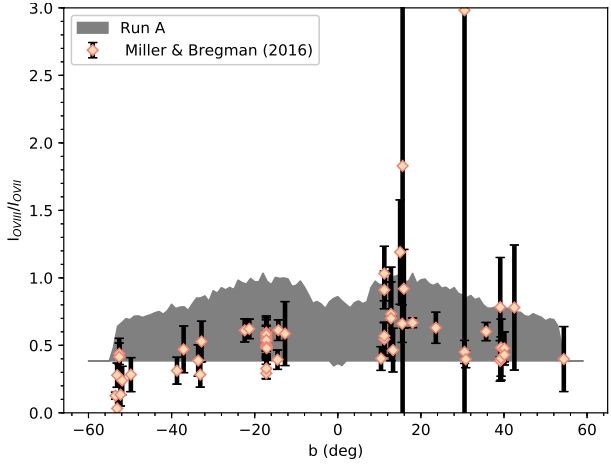}
\caption{Synthetic O VIII to O VII line ratios (shaded area) as a function of Galactic latitude in run A at $t= 5$ Myr. The dots represent the observed O VIII to O VII line ratios adopted from \citet{Miller2016}. The flat lower boundary of the shaded area is mainly contributed by the sight lines that only pass through the unperturbed MW CGM with temperature $0.2$ keV (not passing through the Fermi bubbles). }\label{fig-RunA-o7o8}
 \end{figure}

As implied in Equation \ref{brightness}, the X-ray surface brightness depends on both gas density and temperature. For the hot gas with temperature in the range of $10^{5.5}$--$10^{7}$ K, the emission line strength ratio of O VIII to O VII is a sensitive temperature diagnostic \citep{Miller2016}:
\begin{equation}\label{eq-oviii/ovii}
\frac{I(l,b)_{\tiny{\textrm{OVIII}}}}{I(l,b)_{\textrm{\tiny{OVII}}}}=\frac{\int_{\rm los} n_{H}n_{e}\epsilon(T)_{\textrm{\tiny{OVIII}}}dl}{\int_{\rm los} n_{H}n_{e}\epsilon(T)_{\textrm{\tiny{OVII}}}dl},
\end{equation}
where $\epsilon(T)_{\textrm{\tiny{OVIII}}}$ and $\epsilon(T)_{\textrm{\tiny{OVII}}}$ are the O VIII and O VII line emissivities, respectively. In this subsection, we present the O VIII to O VII line ratios calculated from the simulation data in run A and compare them with observations. 
{ For the initial gas distribution with uniform temperature $T = 0.2$ keV, Equation (\ref{eq-oviii/ovii}) reduces to a constant $I(l,b)_{\tiny{\textrm{OVIII}}}/I(l,b)_{\textrm{\tiny{OVII}}}=\epsilon(T)_{\textrm{\tiny{OVIII}}}/\epsilon(T)_{\textrm{\tiny{OVII}}}$, corresponding to the flat lower boundary of the shaded area in Figure \ref{fig-RunA-o7o8}.}

Figure \ref{fig-RunA-o7o8} shows the simulated O VIII /O VII ratios as a function of Galactic latitude in run A at $t= 5$ Myr. { The shaded area corresponds to the simulated O VIII /O VII ratios covering the full longitude range $0^{\circ} \leq l \leq 360^{\circ}$ for any specific latitude. For a specific sight line, the O VIII /O VII ratio depends on the temperature of the Fermi bubble and its contribution to the X-ray emission along this sight line. The flat lower boundary of the shaded area corresponds to the sight lines that only pass through the unperturbed halo gas with $T=0.2$ keV (not passing through the Fermi bubbles). If the shaded area covers a large fraction of the data points, we conclude that the model is roughly consistent with the observed O VIII /O VII ratios. If most of the data points lie outside of the shaded area, we conclude that the model is not consistent with the observed O VIII /O VII ratios. Some data points at $b\sim -50^{\circ}$ lie substantially below the shaded area, suggesting that most of the hot gas along these sight lines has temperatures lower than our adopted initial gas temperature $0.2$ keV.

As can be seen, the simulated O VIII to O VII line ratios (shaded area) in run A covers a large fraction of the observed line ratios.} This is not surprising, as in this run, the temperature of the shock-compressed hot gas in the Fermi bubbles is about $0.4$ keV (see Fig. \ref{fig-denT1d}), consistent with the value inferred from the observed O VIII /O VII ratios in \citet{Miller2016}. At each Galactic latitude, the variation in the O VIII /O VII ratio is mainly caused by the variation in the contribution of the X-ray flux of the Fermi bubbles to the total X-ray flux along different sight lines. The maximum O VIII /O VII ratio at each latitude tends to decrease with Galactic latitude at $b \gtrsim 15^{\circ}$, which is also due to the decrease of the X-ray flux of the Fermi bubbles with latitude.

\begin{figure}[h!]
 \centering
 \includegraphics[width=8cm]{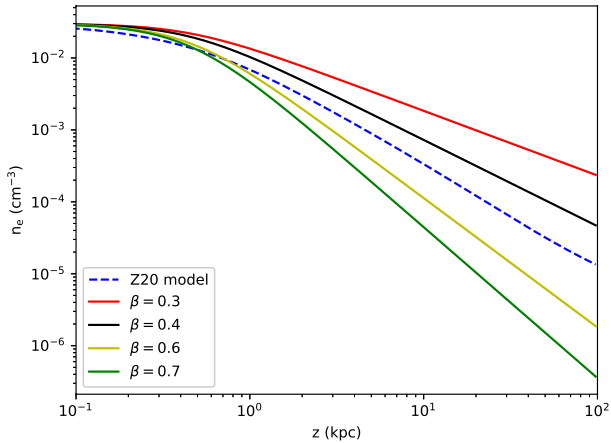}\\
 \caption{Axisymmetric $\beta$ models used as the initial CGM density distributions in runs G, H, I, and J with $\beta=0.3,0.4,0.6,0.7$, respectively. Thermal electron number density is shown as a function of $z$ along the MW rotation axis. The dashed line represents the hydrostatic Z20 model used in run A, which can be approximated by the $\beta$ model with $\beta=0.5$. }\label{fig-den-beta1d}
\end{figure}

 \subsection{The Slope of the CGM Density Profile}\label{sec-dis-slope}

In this subsection, we investigate the slope of the unperturbed gas density profile in the inner Galaxy. As shown in \citetalias{Zhang2020}, the hydrostatic Z20 model used in run A roughly matches the best-fit $\beta$-model in \citet{Miller2015} with $\beta=0.5$ at $r\gtrsim 2$ kpc, which is a power law in radius $n_{\rm e} \propto r^{-1.5}$. More precisely, the Z20 model can be approximated by an axisymmetric $\beta$ model (Equation \ref{beta3}) with $n_{0}=0.03$ cm$^{-3}$, $\beta=0.5$, $R_{c}=0.58$ kpc and $z_{c}=0.45$ kpc. Here, we use this axisymmetric $\beta$ model and vary the value of $\beta$ to investigate the slope of the unperturbed CGM density profile.

\begin{figure}[h!]
 \centering
 \includegraphics[width=8cm]{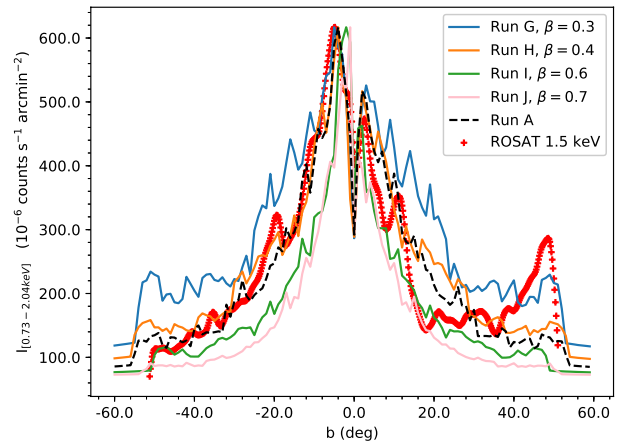}\\
 \caption{Synthetic $0.73-2.04$ keV X-ray surface brightness profiles as a function of Galactic latitude along $l=5^{\circ}$ in runs A, G, H, I, and J. As in Fig. \ref{fig-x-RunA1d}, each line has been rescaled by a factor of $f^{2}$ to better fit the ROSAT data (red pluses), with the corresponding value of $f$ listed in Table \ref{tab-para}.}\label{fig-x-beta1d}
\end{figure}

Runs G, H, I, and J are four representative simulations with $\beta=0.3,0.4,0.6,0.7$, respectively (see Table \ref{tab-para}). The initial CGM density profiles of these four simulations are shown in Figure \ref{fig-den-beta1d}. We calculate the $0.73-2.04$ keV X-ray surface brightness distributions in these runs at $t=5$ Myr with the same method used for run A (see Sec. \ref{sec-x}). The rescaled synthetic $0.73-2.04$ keV surface brightness profile along $l=5^{\circ}$ is shown in Figure \ref{fig-x-beta1d}. Here we focus on the southern hemisphere ($b<0^{\circ}$), as the northern hemisphere is subject to significant additional X-ray emissions and absorptions.

\begin{deluxetable*}{ccccccccccc}\label{tab-static}
\tablecaption{Several Key Parameters of the Simulated Fermi Bubbles in Our Simulations}
\tablewidth{0pt}
\tablehead{{Run} &\colhead{A}&\colhead{B}&\colhead{C}&\colhead{D}&\colhead{E}&\colhead{F}&\colhead{G}&\colhead{H}&\colhead{I}&\colhead{J}
\\
\colhead{CGM model} &Z20&MB&G20&NFW&L17&S19&$\beta=0.3$&$\beta=0.4$&$\beta=0.6$&$\beta=0.7$}
\startdata
EMD &0.15&5.05&3.91&1.09&0.26&0.99&1.51&0.77&0.88&1.58\\
$w$&0.81&0.97&0.98&0.77&0.97&0.67&0.95&0.91&0.89&0.86\\
\enddata
\tablecomments{$w$ is the conical parameter defined as the ratio between the bubble width at $b=15^{\circ}$ and that at $b=30^{\circ}$. For the observed northern and southern gamma-ray bubbles, $w\simeq 0.83$, $0.75$, respectively. EMD is the Earth Mover's Distance, a measure of the distance between the observed and simulated 0.73-2.04 keV X-ray surface brightness profiles as a function of Galactic latitude along $l=5^{\circ}$. Here we calculate the values of the EMD for the regions with $-40^{\circ}<b<0^{\circ}$. A lower EMD value corresponds to a better fit between the model prediction and the data.}
\end{deluxetable*}

{ As shown in Figure \ref{fig-x-beta1d} ($b<0^{\circ}$), the X-ray surface brightness profile is a sensitive diagnostic of the unperturbed CGM density slope in the inner Galaxy. It appears that run A ($\beta \approx 0.5$) provides the best fit to the ROSAT X-ray surface brightness profile, while run H ($\beta=0.4$) also results in a reasonably good fit to the data. To be more quantitative, we adopt a statistical tool --- the Earth Mover's Distance (EMD; see \citealt{Rubner2000}), which provides a measure of the distance between the observed and simulated X-ray surface brightness profiles shown in Figure \ref{fig-x-beta1d}. We calculate the EMDs of our simulated X-ray surface brightness profiles with respect to the ROSAT data for the southern hemisphere at Galactic latitudes $-40^{\circ}<b<0^{\circ}$. At $b<-40^{\circ}$, the bubble emission is quite low, while the X-ray emissions from other sources become important. The calculated values of the EMDs for our runs are shown in Table \ref{tab-static}. Higher EMD values corresponds to larger differences between the model predictions and the observational curve. As can be seen in Table \ref{tab-static}, run A ($\beta \approx 0.5$) has the lowest EMD value of $0.15$, confirming that run A  ($\beta \approx 0.5$) provides the best fit to the observed X-ray surface brightness profile. At $r\gtrsim 1$ kpc, the unperturbed CGM density profile with $\beta=0.5$ roughly scales as $n_{\rm e} \propto r^{-1.5}$.}

\subsection{Other CGM Density Models}

In this subsection, we present the simulation results of the other CGM density models --- the MB, G20, NFW, L17, and L19 models in runs B, C, D, E, F, respectively (see Table \ref{tab-para}). As shown in Figure \ref{fig-den1d}, these density profiles differ significantly in the inner Galaxy. The MB and G20 models are quite flat in the inner Galaxy, while the cuspy NFW density model rises exponentially toward the GC. In the disk-like L17 and S19 models, the hot gas density decreases exponentially away from the Galactic plane. As described in detail below, in addition to the X-ray surface brightness distribution, the morphology of the simulated Fermi bubble (forward shock) is also significantly affected by the jet properties and the ambient CGM density distribution \citep[see also][]{Sofue2019}.

{ The Fermi bubbles have a unique morphology. Both the X-ray and gamma-ray images of the Fermi bubbles indicate that the bubble width gradually increases with latitude at low latitudes with $|b| \lesssim 30^{\circ}$. To quantitatively compare the simulated bubble morphology with observations, here we introduce a conical parameter $w$, which is defined as the ratio between the bubble width at $b=15^{\circ}$ ($-15^{\circ}$) and that at $b=30^{\circ}$ ($-30^{\circ}$). We choose these two representative latitudes because at $b\sim15^{\circ}$ the bubble region is significantly affected by the gamma ray emissions from the Galactic plane, and at $b\sim30^{\circ}$ the observed gamma-ray bubble reaches its maximum width. $w=1$ for an idealized cylindrical bubble. $w \simeq0.83$, $0.75$ for the observed northern and southern gamma-ray Fermi bubble, respectively. As can be seen in Table \ref{tab-static}, run A ($w=0.81$) and run D ($w=0.77$) are much more consistent with observations than the other runs.}

\subsubsection{ Flat Density Models: MB and G20}\label{sec-flat}

\begin{figure*}
\gridline{
     \fig{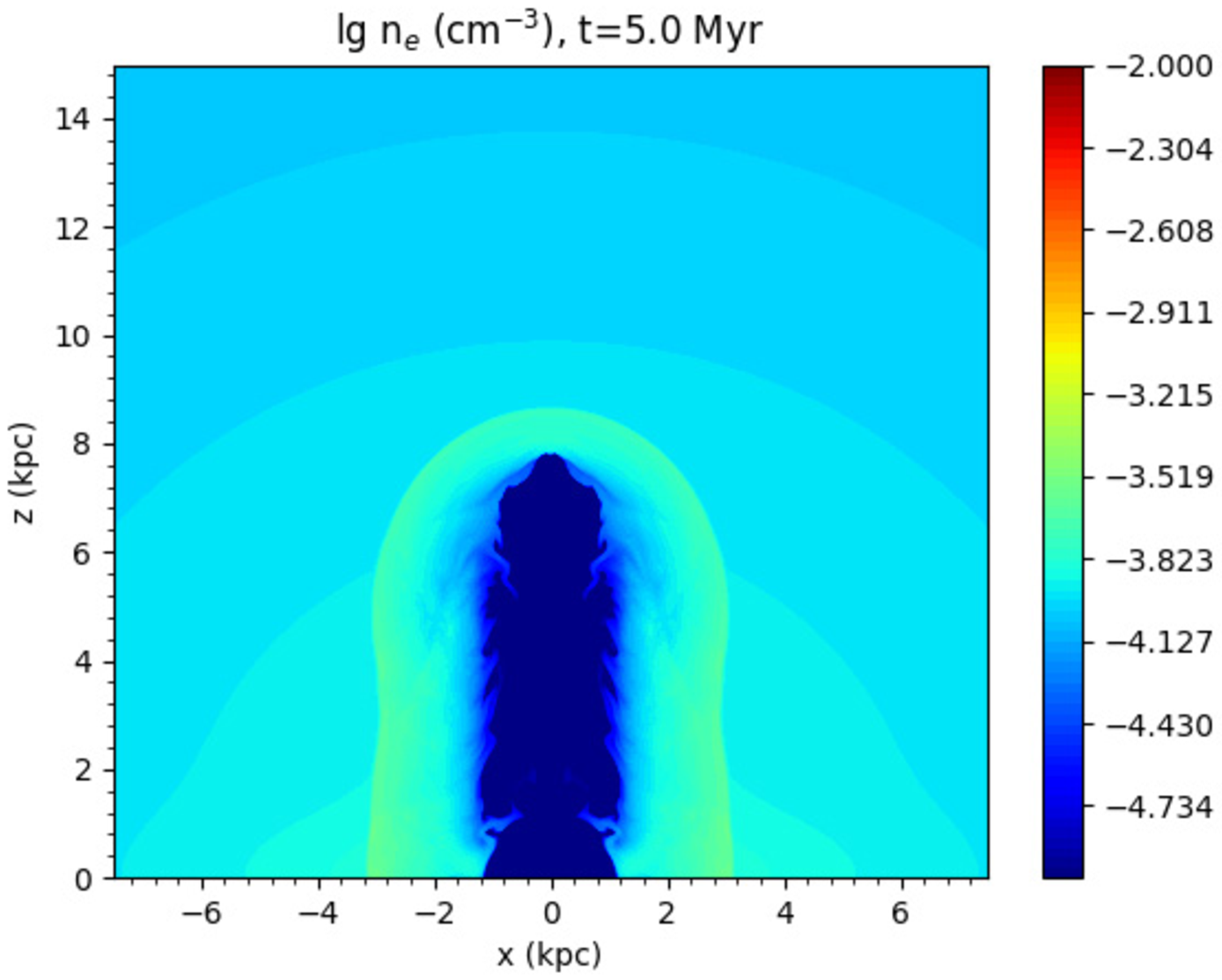}{0.25\textwidth}{}
     \fig{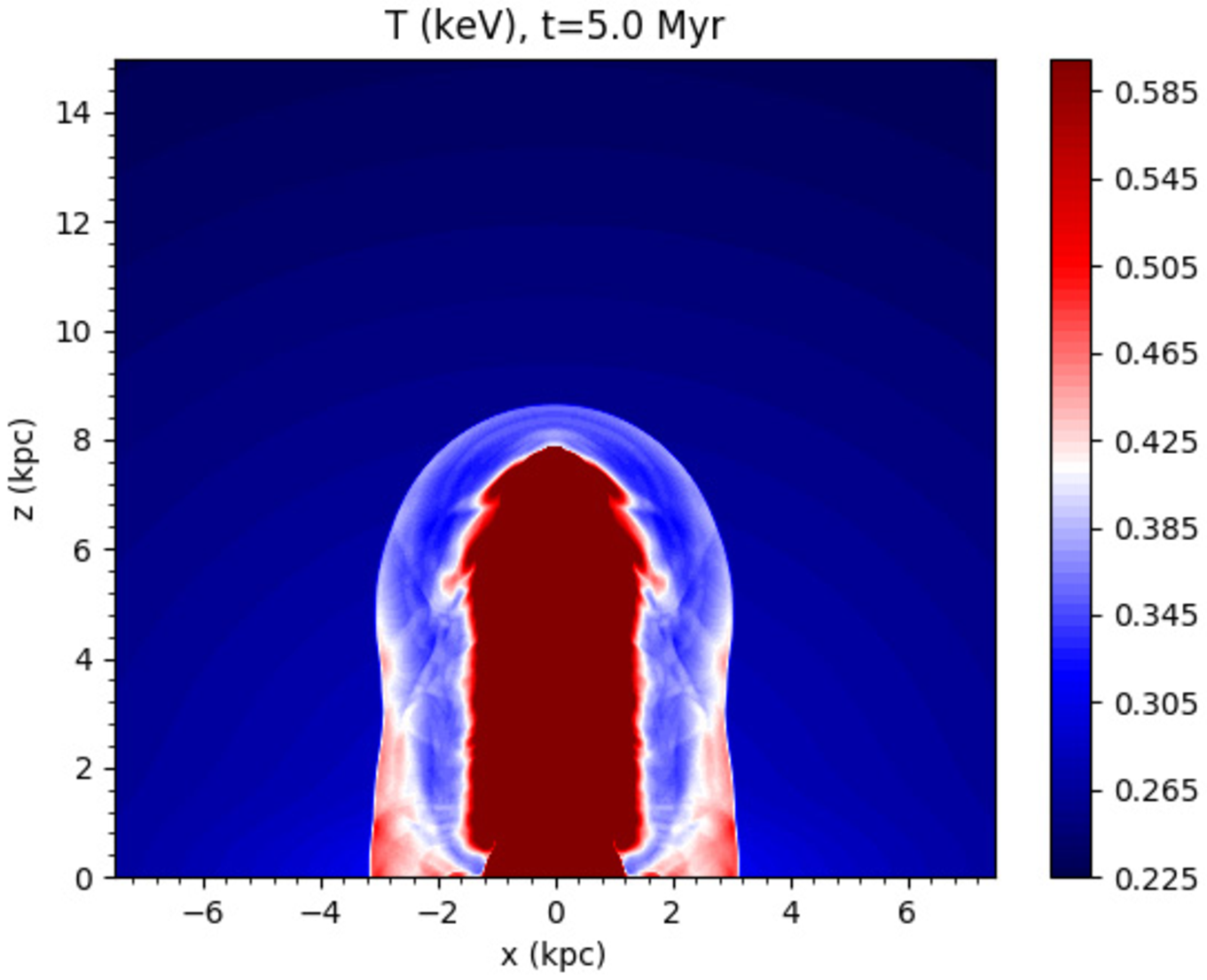}{0.25\textwidth}{}
     \fig{x-mb.eps}{0.25\textwidth}{}
     \fig{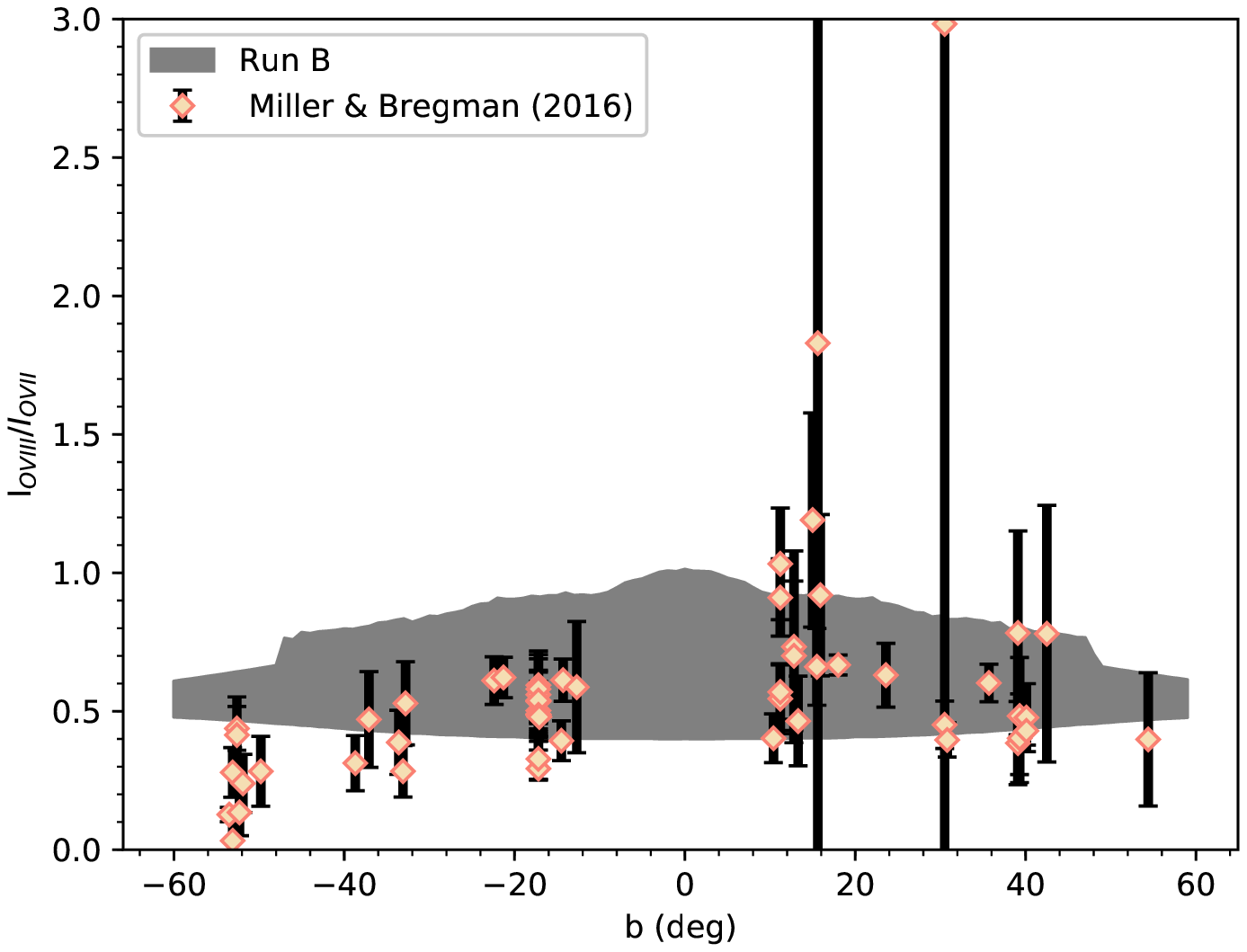}{0.25\textwidth}{}
     }\vspace{-24pt}
\gridline{
     \fig{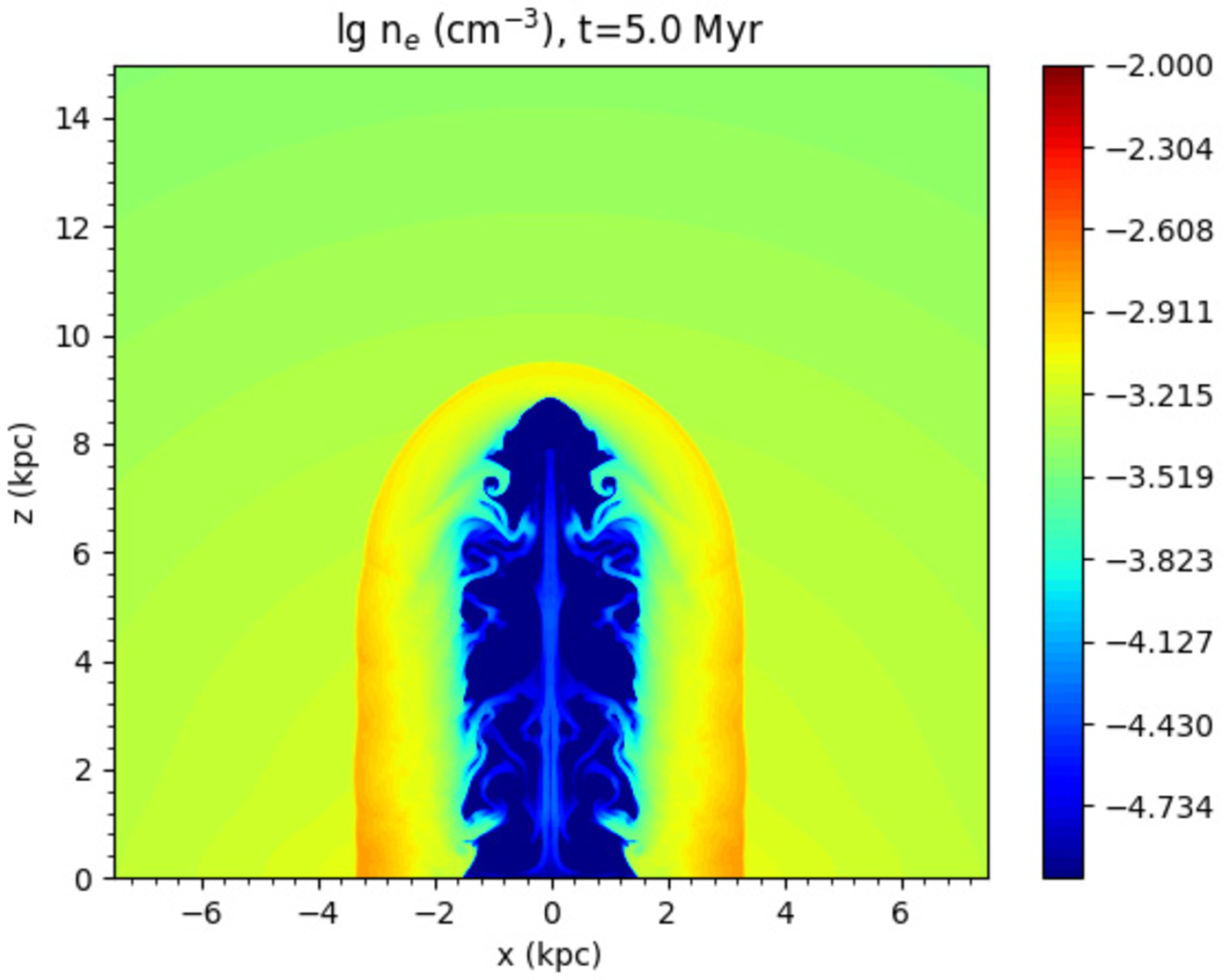}{0.25\textwidth}{}
     \fig{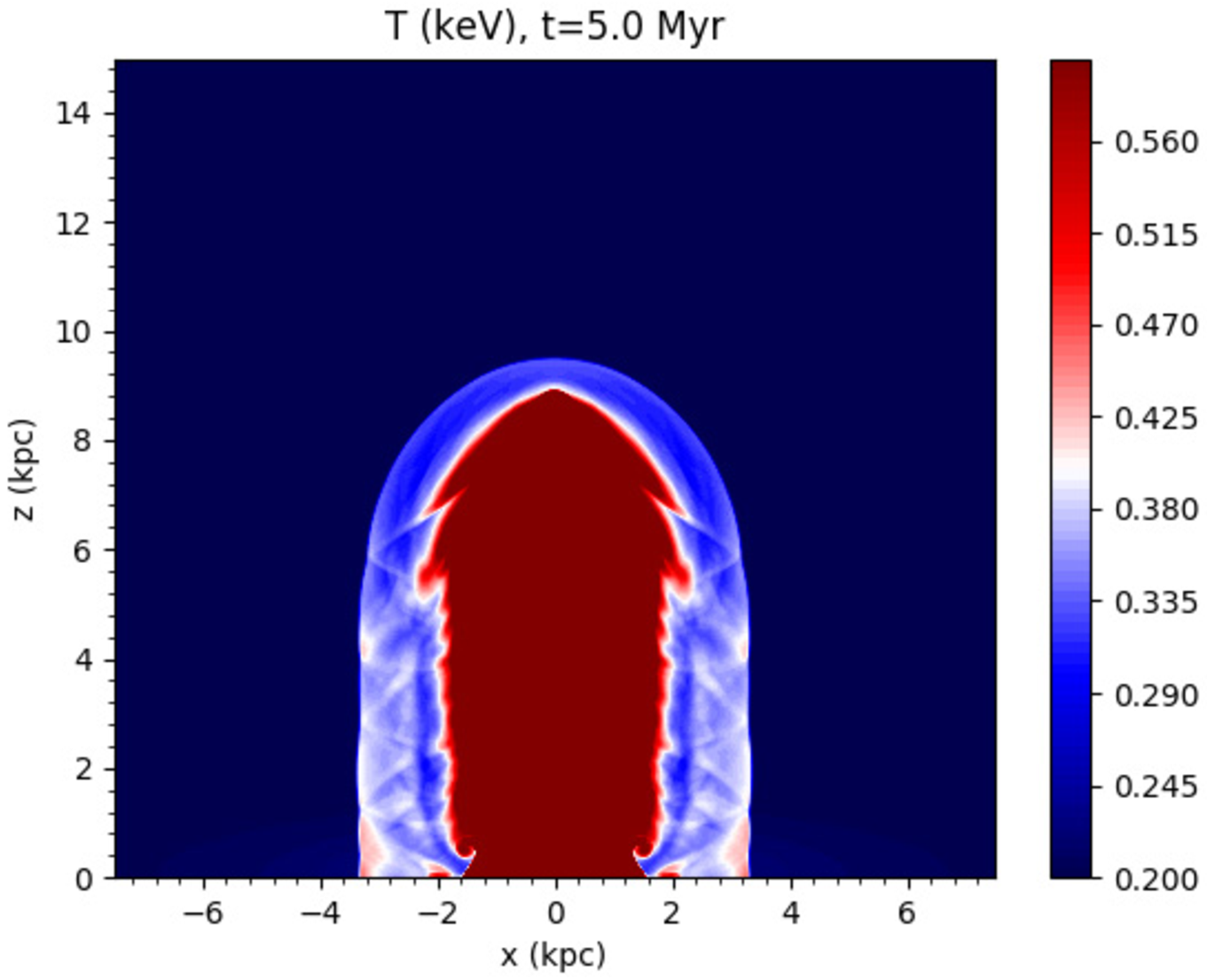}{0.25\textwidth}{}
     \fig{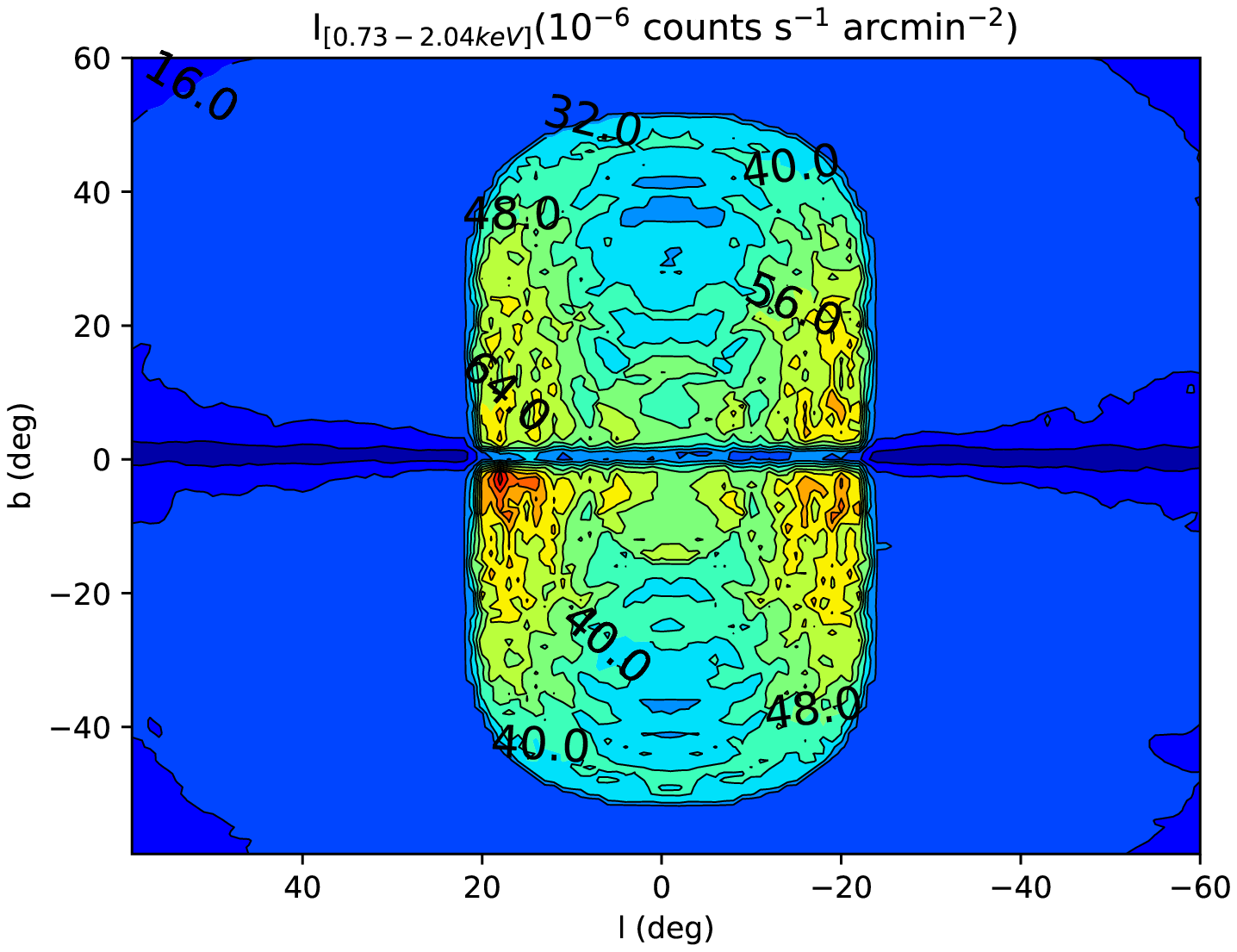}{0.25\textwidth}{}
     \fig{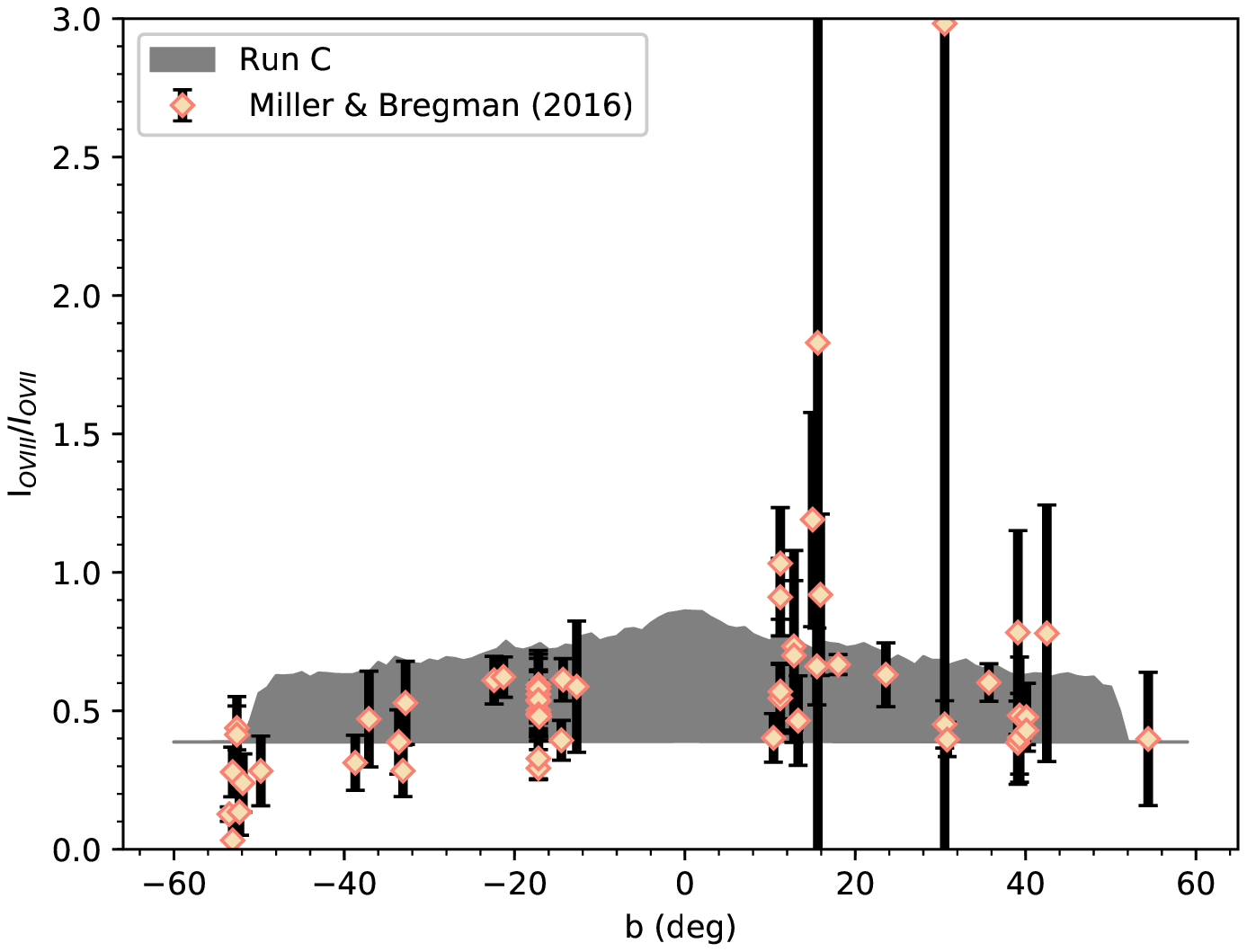}{0.25\textwidth}{}
     }\vspace{-24pt}
\gridline{
     \fig{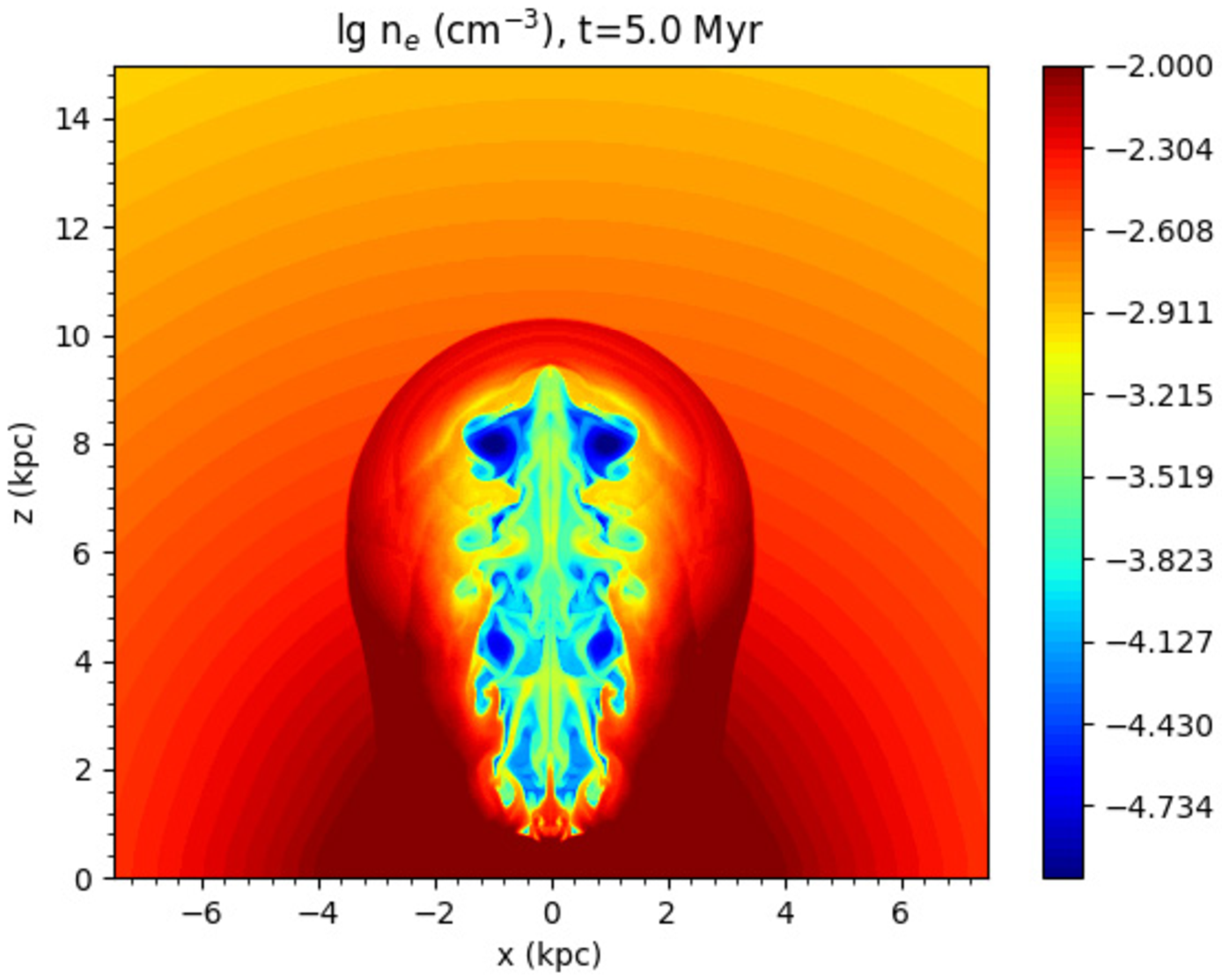}{0.25\textwidth}{}
     \fig{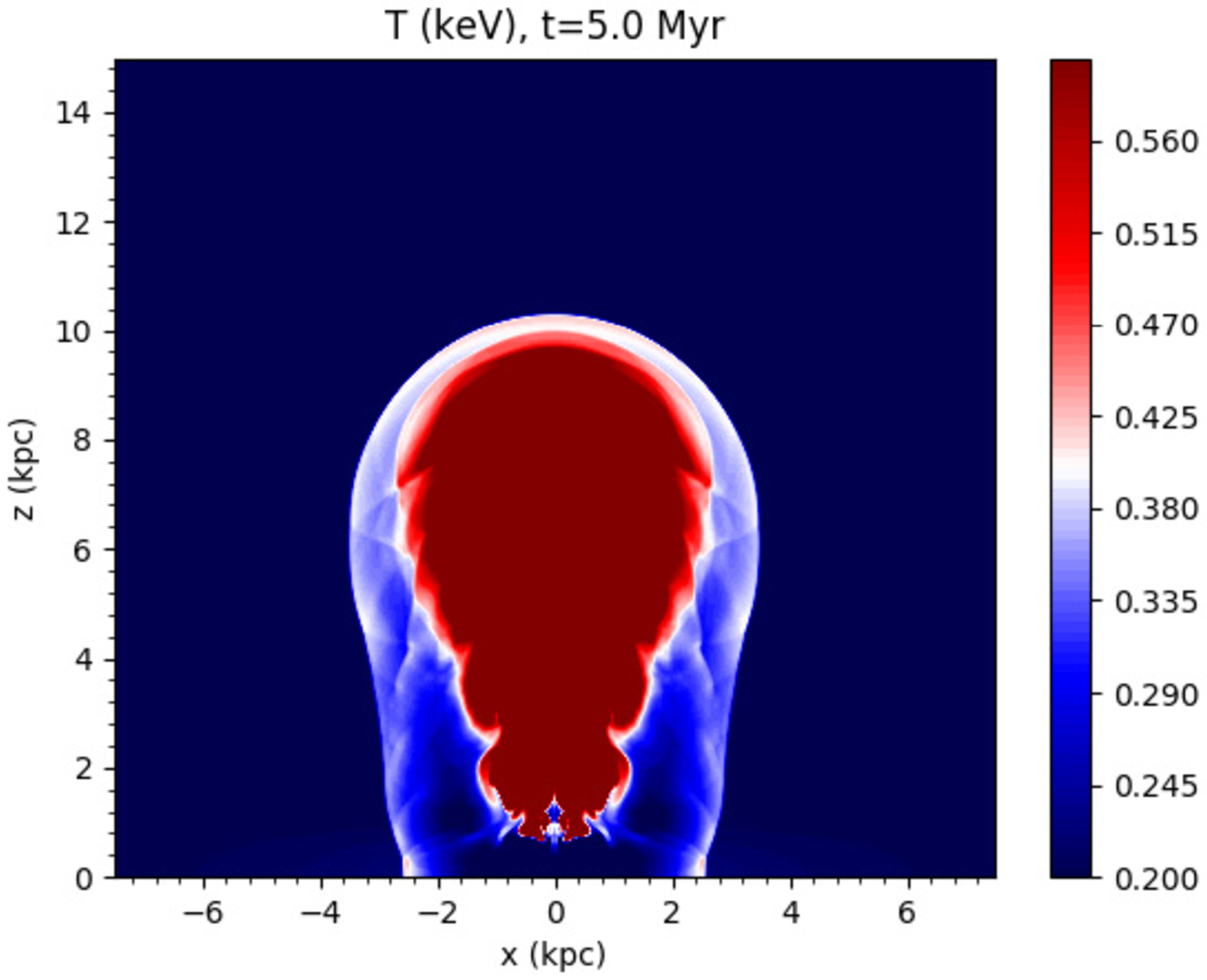}{0.25\textwidth}{}
     \fig{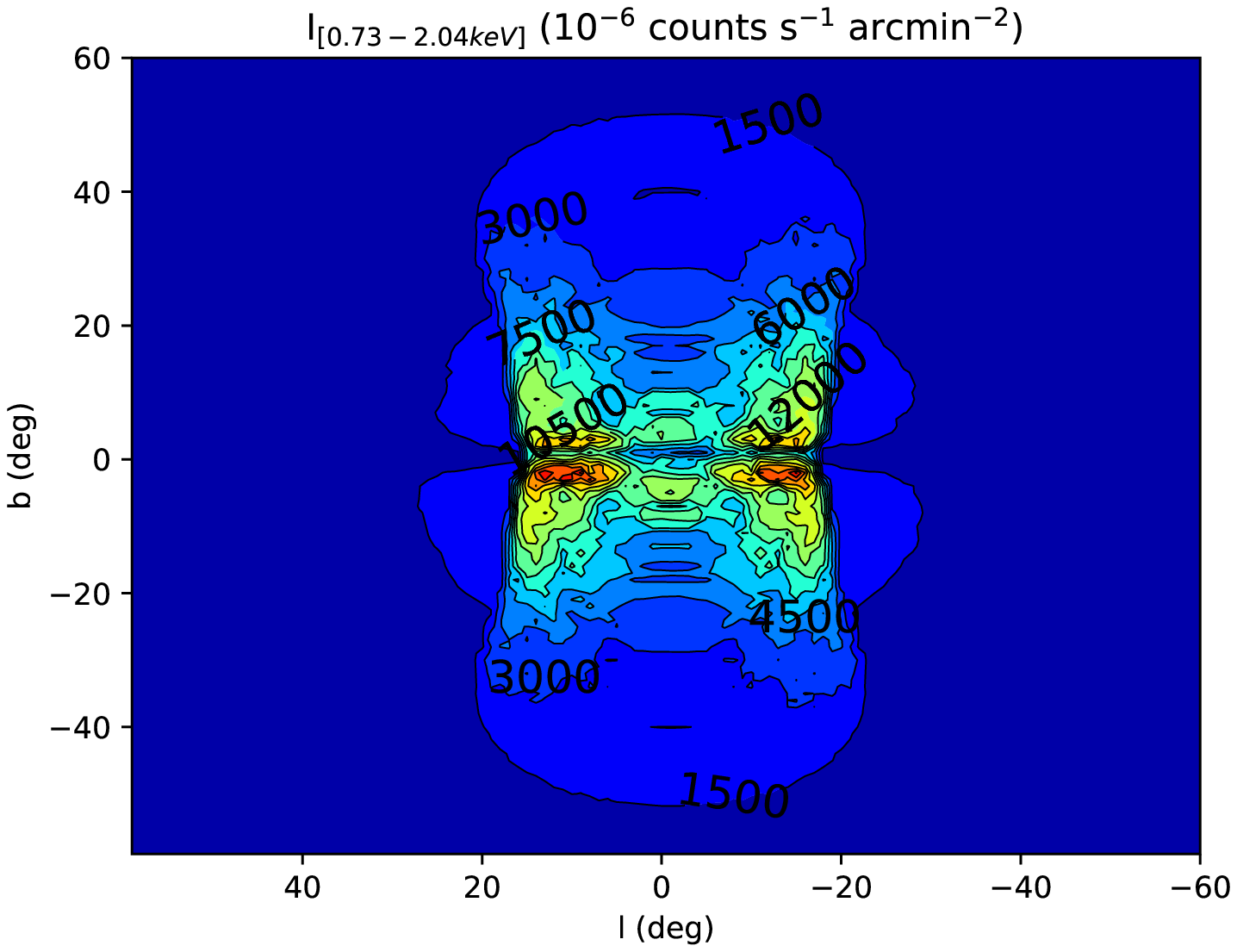}{0.25\textwidth}{}
     \fig{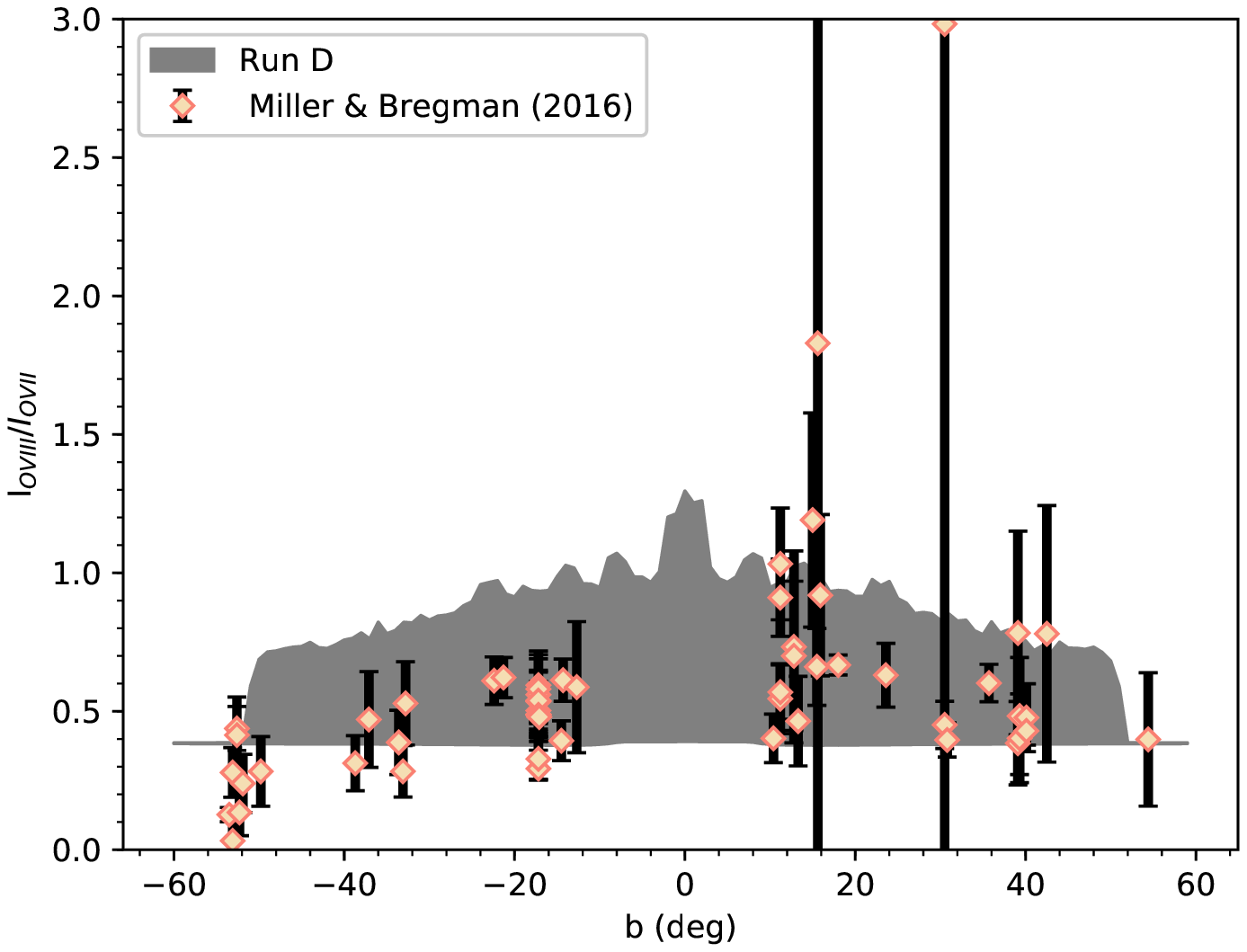}{0.25\textwidth}{}
     }
\vspace{-24pt}
\gridline{
     \fig{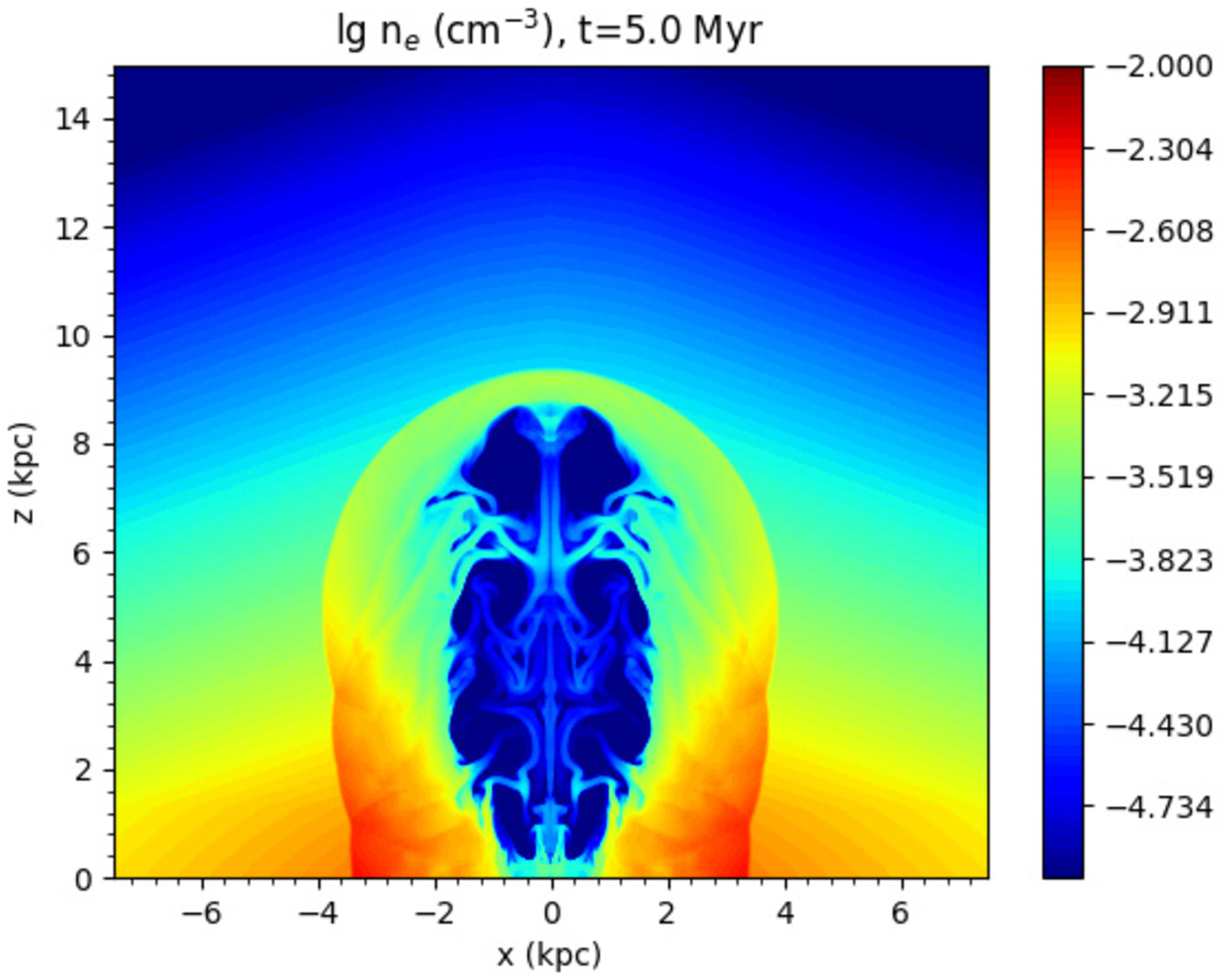}{0.25\textwidth}{}
     \fig{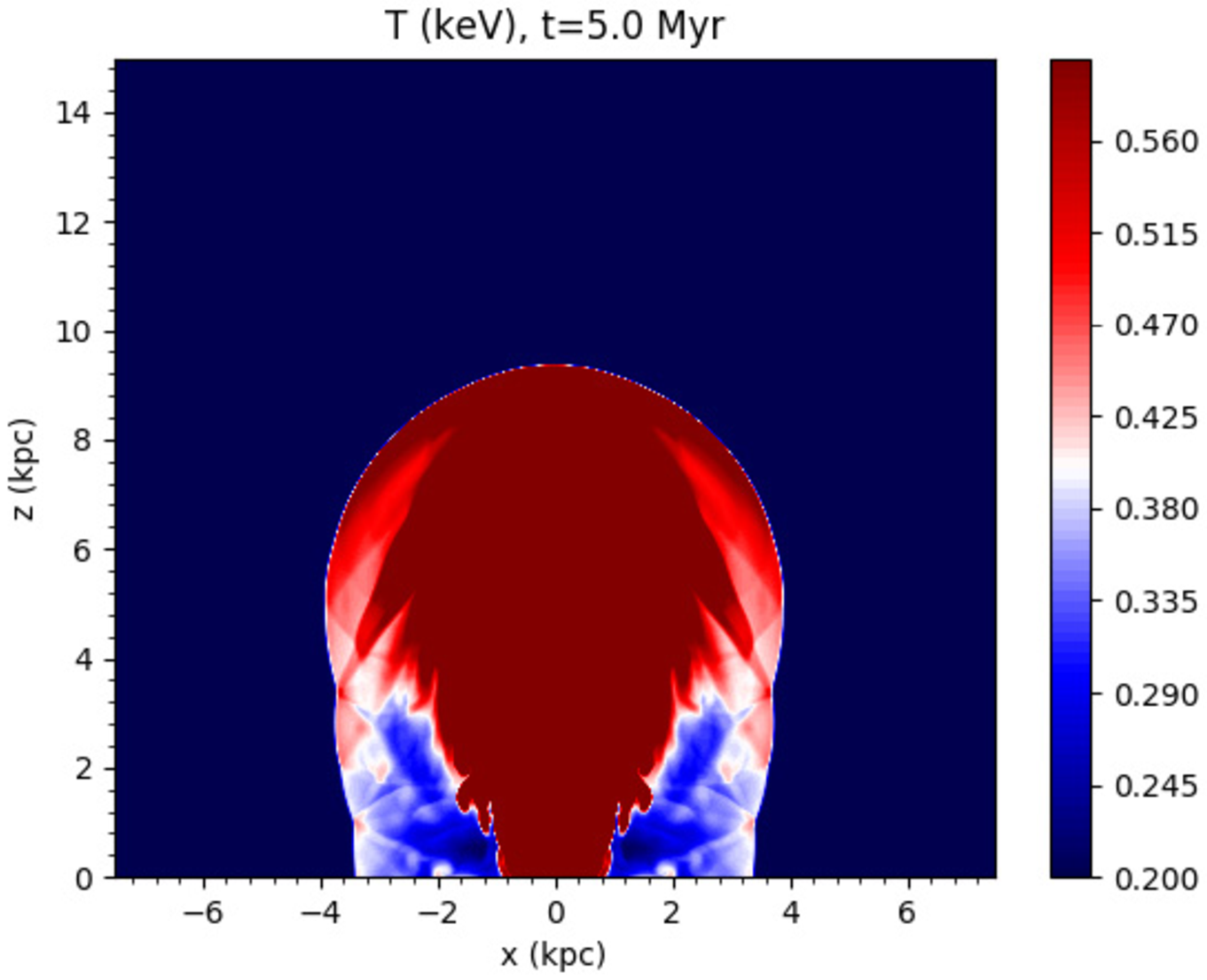}{0.25\textwidth}{}
     \fig{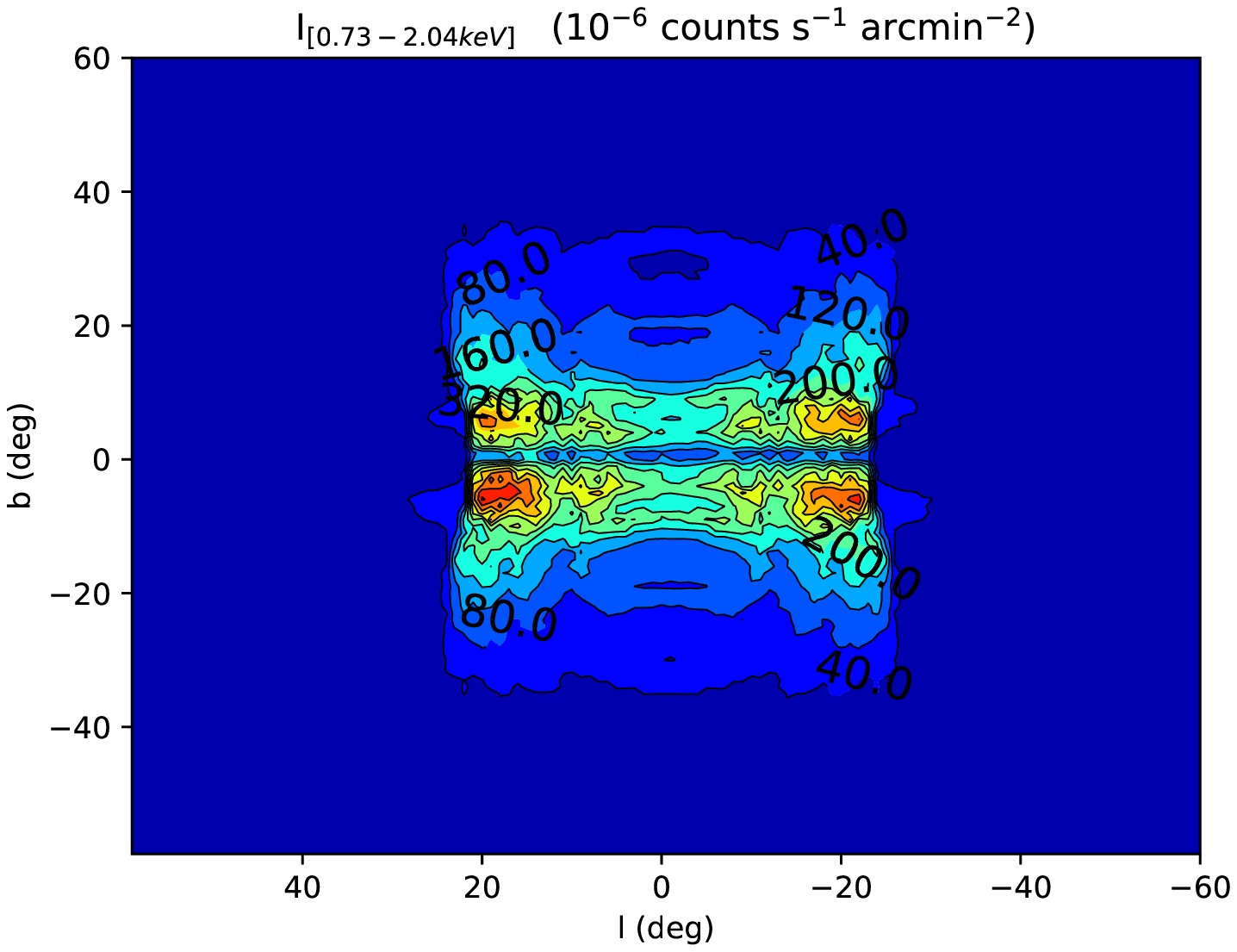}{0.25\textwidth}{}
     \fig{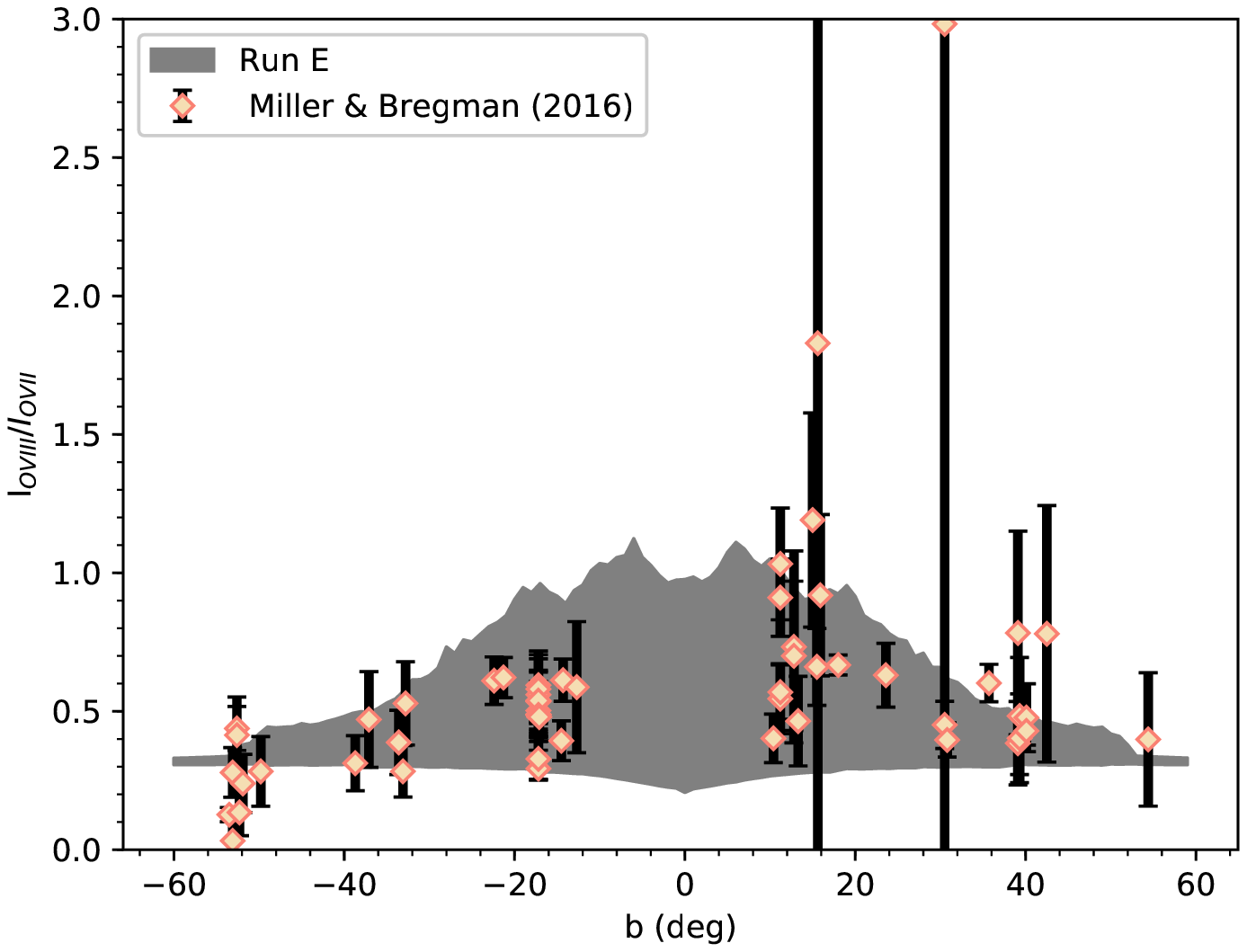}{0.25\textwidth}{}
     }
     \vspace{-24pt}
\gridline{
     \fig{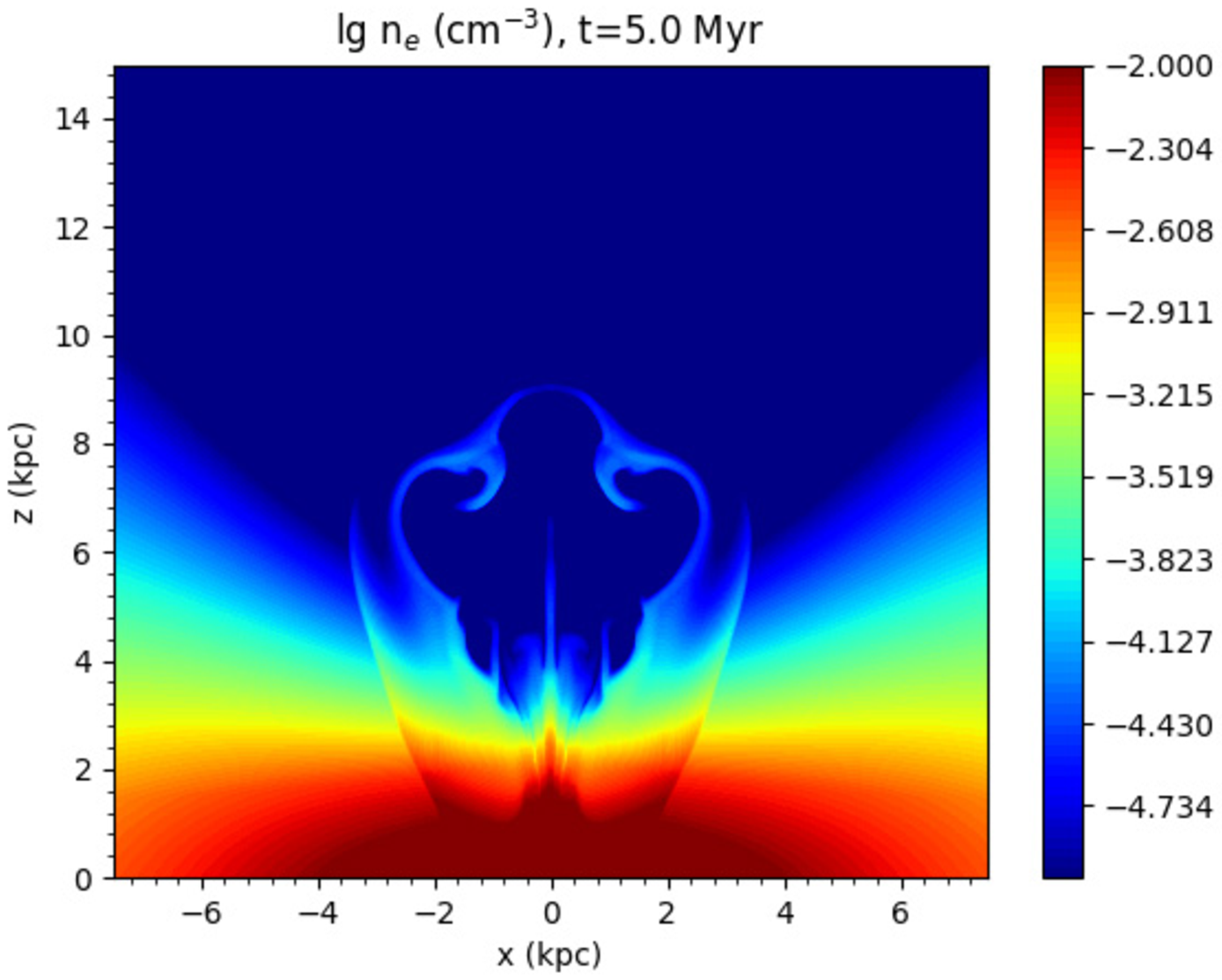}{0.25\textwidth}{}
     \fig{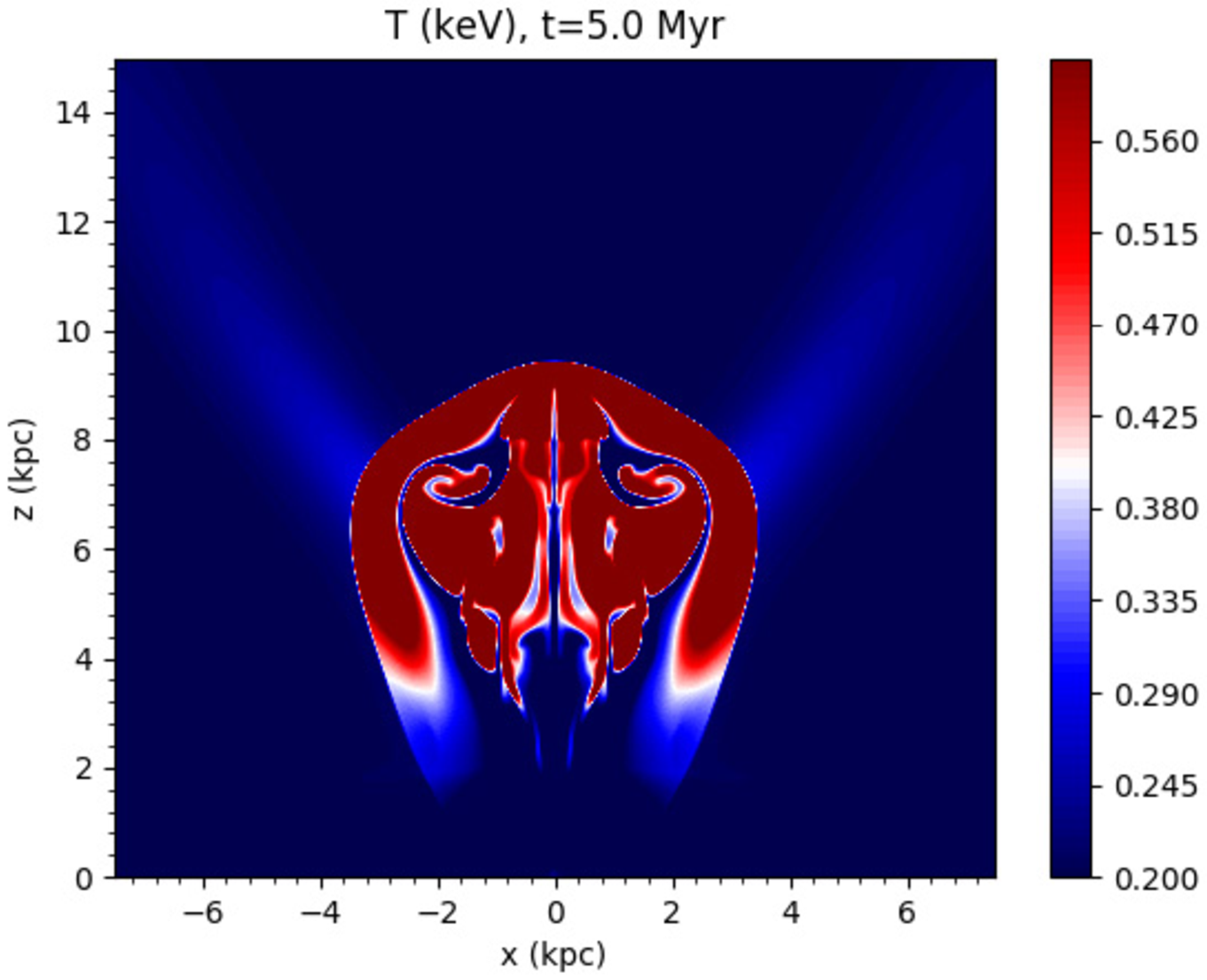}{0.25\textwidth}{}
     \fig{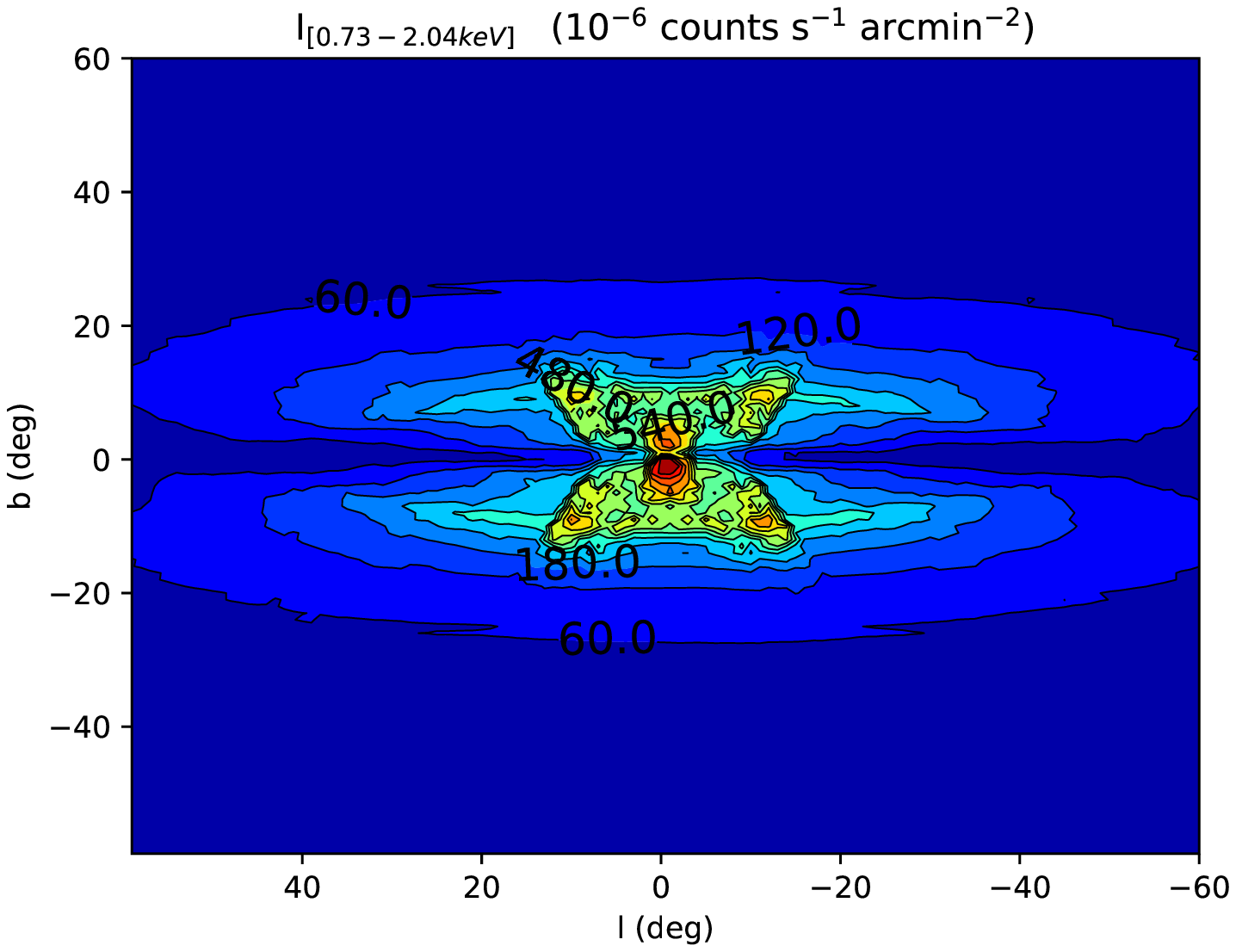}{0.25\textwidth}{}
      \fig{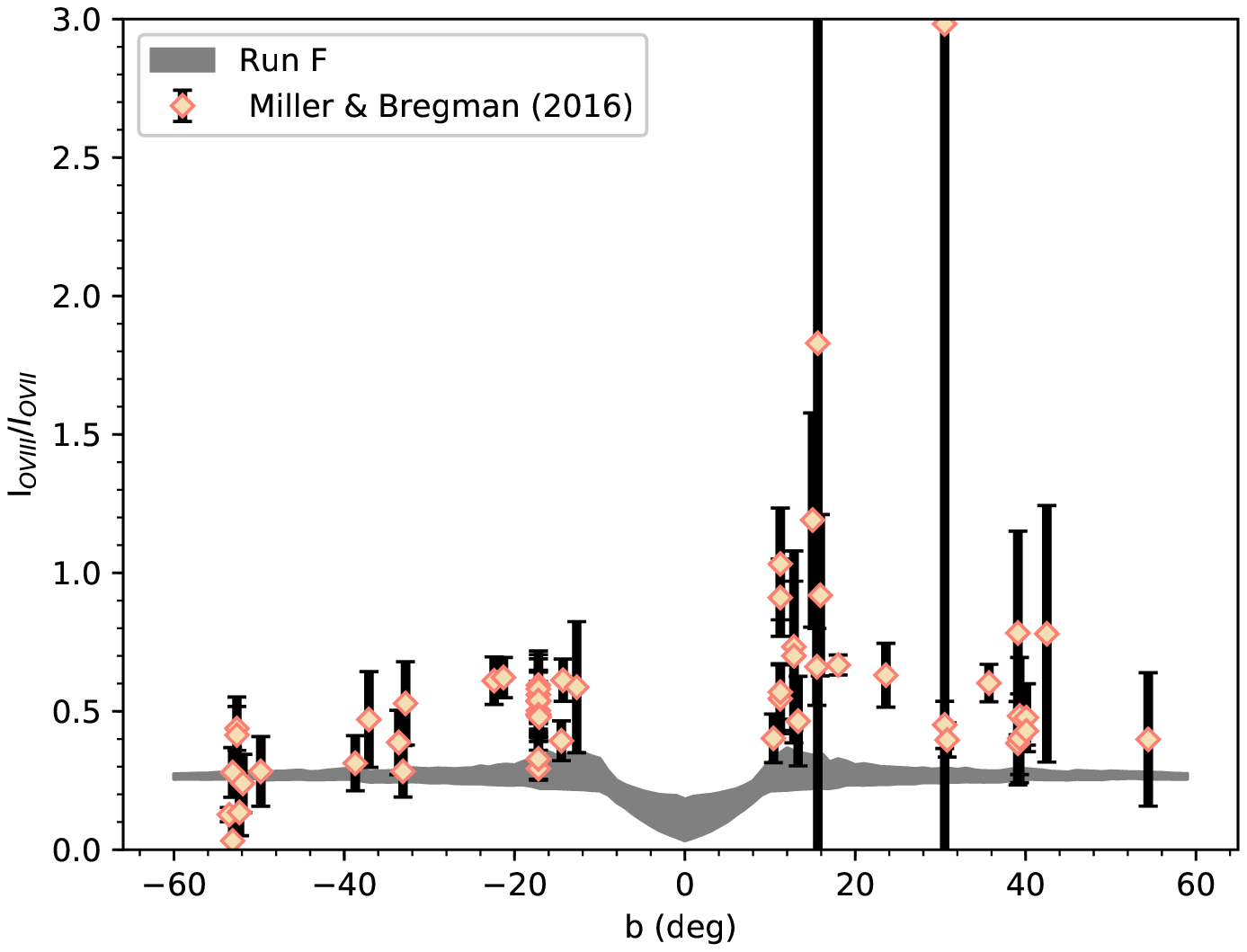}{0.25\textwidth}{}
     }
\caption{Simulation results at $t=5$ Myr in runs B, C, D, E, and F, shown from the top row to the bottom row. Each row has four panels, which from left to right, represent the distributions of the gas density, temperature, the synthetic $0.73-2.04$ keV X-ray surface brightness, and the O VIII/O VII emission line ratio.}\label{fig-total0}
\end{figure*}

Runs B and C simulate the formation of the Fermi bubbles in the MB and G20 models, respectively. In both models, the CGM density profiles in the inner Galaxy are relatively flat (Fig. \ref{fig-den1d}). As can be seen in the top two rows of Figure \ref{fig-total0}, the morphologies of the resulted forward shocks in these two runs are nearly cylindrical { with $w\approx 1$ (see Table \ref{tab-static}), and the bubble bases near the Galactic plane are very wide}.  We have experimented with a large number of simulations and found that the cylindrical Fermi bubbles are a generic result of the MB and G20 models, which is inconsistent with the bilobular morphology of the observed Fermi bubbles with narrow bases. Comparing to flat CGM density profiles, a radially-decreasing density profile as in the Z20 model allows the jet to propagate faster, and deposit energy to large distances much more quickly, leading to narrow bases of the shock front near the GC as observed.

\begin{figure}[h!]
 \centering
 \includegraphics[width=8cm]{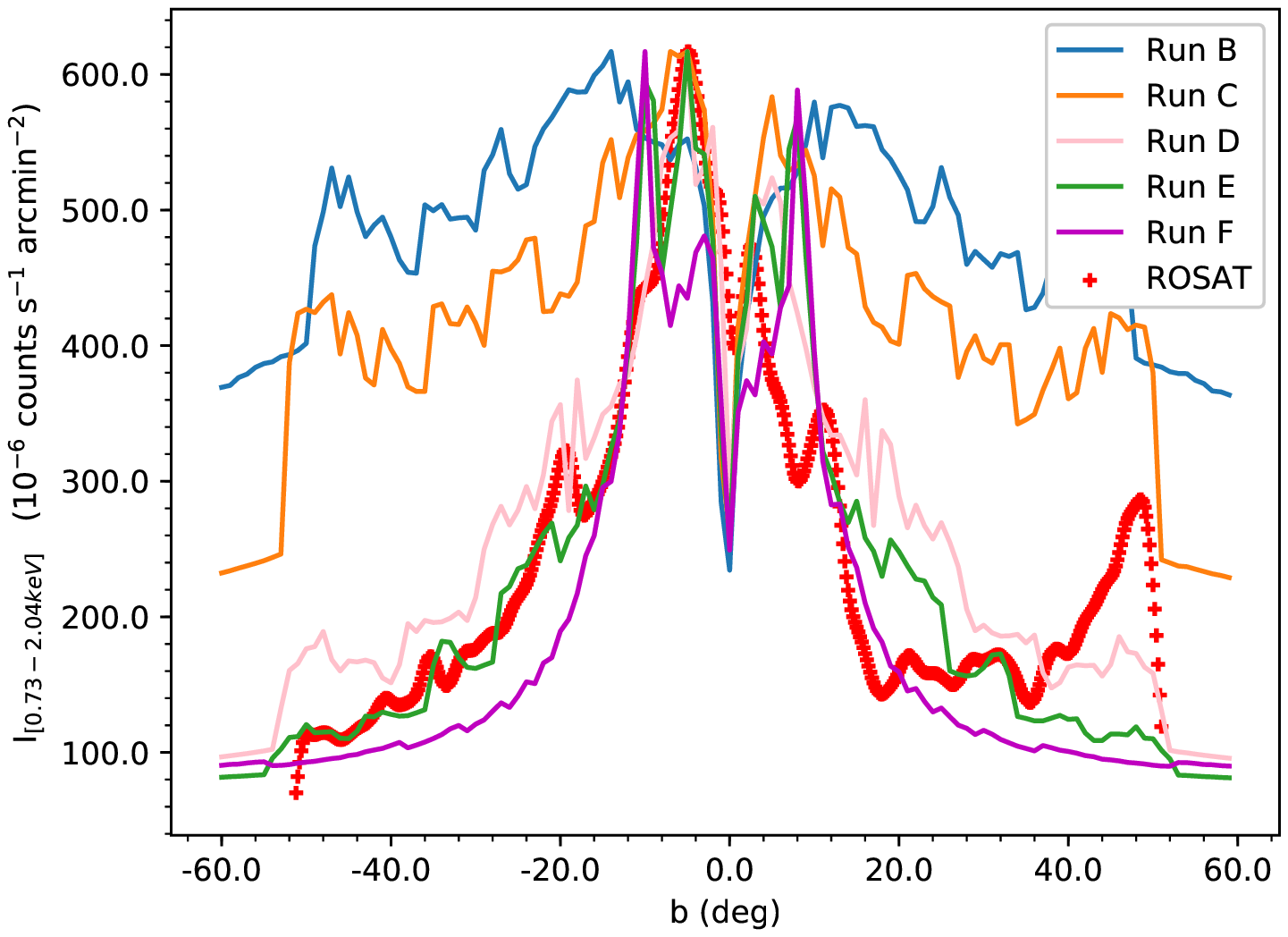}
\caption{Synthetic $0.73-2.04$ keV X-ray surface brightness profiles as a function of Galactic latitude along $l=5^{\circ}$ in runs B, C, D, E, and F. Each line has been rescaled by a factor of $f^{2}$ to better fit the ROSAT data, with the corresponding value of $f$ listed in Table \ref{tab-para}. The CGM density models adopted in runs B--F are the MB, G20, NFW, L17, and S19 models, respectively. }\label{fig-x-all-1d}
 \end{figure}

The synthetic X-ray surface brightness distributions in runs B and C are also inconsistent with ROSAT observations. As shown in Figure \ref{fig-total0}, the predicted X-ray surface brightnesses of the Fermi bubbles in runs B and C are about one-two orders of magnitude lower than the observed value, indicating that the real hot gas density in the inner Galaxy is higher than those in the MB and G20 models. This discrepancy could be resolved by rescaling the initial CGM density profile with a factor $f$ as described in Sec. \ref{sec-x}. The resulted X-ray surface brightness profiles from runs B and C are shown as a function of Galactic latitude along $l=5^{\circ}$ in Figure \ref{fig-x-all-1d}, and are clearly much flatter than the observed X-ray surface brightness profile by ROSAT. 
{ As shown in Table \ref{tab-static}, the EMD values in runs B and C are more than one order of magnitude higher than that in run A,}
 indicating that the initial CGM density profile in the inner Galaxy should decrease with radius much faster than in the flat MB and G20 models.

\subsubsection{The Cuspy NFW Model}

Run D explores the cuspy NFW model, in which the hot gas density profile rises exponentially toward the GC as $n_{\rm e} \propto r^{-1}$. As shown in the third row of Figure \ref{fig-total0}, the morphology, temperature, and O VIII/O VII ratio of the resulted Fermi bubble are all roughly consistent with observations. The major discrepancy is that the X-ray surface brightnesses predicted by this model is more than one order of magnitude higher than the observed values. As for other models, this discrepancy could be resolved by rescaling the initial CGM density profile with a factor $f$ as described in Sec. \ref{sec-x}. Figure \ref{fig-x-all-1d} shows the rescaled X-ray surface brightness profile from run D as a function of Galactic latitude along $l=5^{\circ}$. As clearly shown, the cuspy NFW model performs much better than the flat MB and G20 models, but the predicted X-ray surface brightnesses at $|b|\gtrsim 15^{\circ}$ are still significantly higher than the observed values, which is expected as in the inner Galaxy, the NFW profile ($\propto r^{-1}$) is significantly flatter than our best-fit Z20 model ($\propto r^{-1.5}$). { It can also be seen in Table \ref{tab-static} that the EMD value in run D is significantly higher than that in run A, although it is much lower than the EMD values in runs B and C.}

\begin{figure}[h!]
 \centering
 \includegraphics[width=8cm]{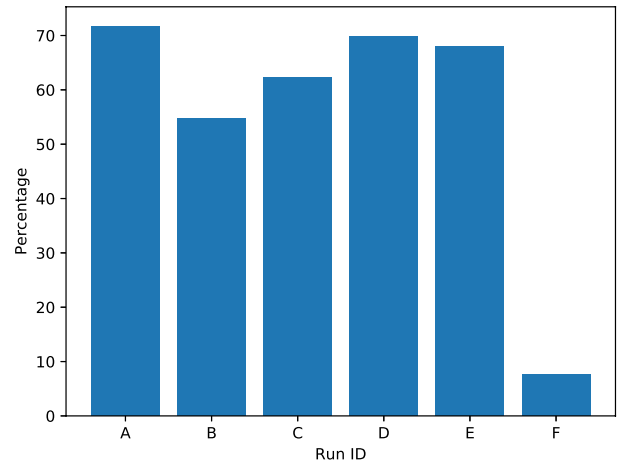}\\
 \caption{Fraction of the observed O VIII to O VII line ratios of the Fermi bubble regions in \citet{Miller2016} reproduced by our simulations. The fraction is very low in run F ($<10\%$), as most of the observed line ratios lie significantly above the simulated values (shaded area) in the bottom-right panel of Fig. \ref{fig-total0}.}\label{fig-bar}
\end{figure}

\subsubsection{Disk-like Density Models: L17 and S19}

The fourth and fifth rows of Figure \ref{fig-total0} show the simulation results of the disk-like L17 and S19 models in runs E and F, respectively. As shown in Fig. \ref{fig-den1d}, the L17 CGM density profile is very flat along the $z$ direction in the inner $5$ kpc. Similar to the flat MB and G20 models, such a flat CGM density profile leads to the cylindrical Fermi bubble 
{ ($w=0.97 \approx 1$, see Table \ref{tab-static}),}
 inconsistent with the bilobular morphology of the observed Fermi bubbles with narrow bases. In the S19 model, the gas density profile is relatively flat only within $|z|\lesssim 0.5$ kpc, and decreases very fast as the value of $|z|$ further increases. While the resulted Fermi bubble in run F has a very narrow base, the gas densities in the bubble at high latitudes are too low to be consistent with observations. In fact, the gas density in the S19 model quickly drops to $<10^{-5}$ cm $^{-3} $ at $|z|> 3$ kpc. 
{ The simulated X-ray surface brightnesses in run F at $|b|\gtrsim 15^{\circ}$ are dominated by the emissions from the unperturbed halo gas, and are significantly lower than the observed values (Fig. \ref{fig-x-all-1d}). Since the X-ray emission of the resulted low-density Fermi bubble is too low, the simulated O VIII/O VII ratios in run F are dominated by the $0.2$ keV halo gas outside the bubble, and are thus significantly lower than the observed values (the bottom-right panel of Fig. \ref{fig-total0}). To be more quantitative, we show the fractions of the observed O VIII to O VII line ratios \citep{Miller2016} reproduced by our simulations in Figure \ref{fig-bar}. While this fraction is relatively high in runs A-E, it is distinctly low ($<10\%$) in run F.}

Therefore, { combining the bubble morphology, X-ray surface brightness distribution, and O VIII/O VII ratios,} we conclude that the disk-like L17 and S19 models are not a good description of the hot gas density distribution in the inner Galaxy before the Fermi bubble event. We note that our study could not probe the halo gas distribution beyond the inner Galaxy. Even for the inner Galaxy, although our study suggests that disk-like models are disfavored, we have not yet ruled out all disk-like models with a large parameter space.

\section{Conclusion and Discussion}

In this work, we combine hydrodynamic simulations and X-ray observations of the Fermi bubbles to investigate the halo gas distribution in the inner Galaxy before the Fermi bubble event. 
{ We assume that the Fermi bubbles and the biconical X-ray structure at the GC have the same origin, evolved from the forward shock driven by a past AGN jet event (\citetalias{Zhang2020}). Cosmic ray acceleration and the associated non-thermal emissions in the shock scenario have been previously investigated (e.g., \citealt{Fujita2013}, \citealt{Fujita2014}, \citealt{Keshet2017}).}
 We consider a variety of representative spherical and disk-like MW CGM models, and use them as initial conditions in a series of simulations to study the formation of the Fermi bubbles. To constrain the initial CGM distribution, we compare the morphology, temperature, X-ray surface brightness distribution, and O VIII/O VII ratios of the simulated Fermi bubbles with relevant observations.

\begin{figure}[h!]
 \centering
  \includegraphics[width=8cm]{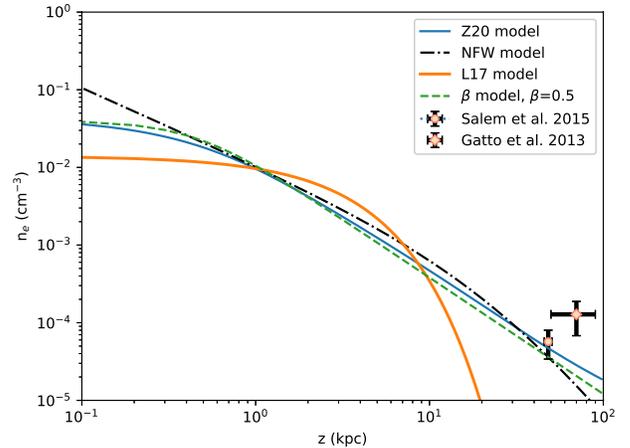}
\caption{Thermal electron number density profiles along the $z$ axis in the Z20, NFW, and L17 models, which are used as initial conditions in runs A, D, and E, respectively. Each line has been rescaled from the original model in Section \ref{sec-models} by a factor of $f$ to better fit the ROSAT data at the current time (see Sec. \ref{sec-x}), with the corresponding value of $f$ listed in Table \ref{tab-para}. The dashed line refers to the $\beta$ model (Eq. \ref{eq-beta-flat-appro}), which provides a reasonably good fit to our best-fit Z20 model in the inner Galaxy. The diamond and circle data points correspond to the density estimates by \citet{Gatto2013} and \citet{Salem2015}, respectively. As in \citet{Guo2020}, the estimated total number densities in these two references have been converted to $n_{\rm e}$ here.}\label{fig-den1d-cali}
 \end{figure}

We find that among our investigated seven CGM models, the best-fit model is the \citetalias{Zhang2020} model, which can be approximated by a nearly-spherical $\beta$ model with $\beta=0.5$ (see Fig. \ref{fig-den1d-cali}):
\begin{equation}\label{eq-beta-flat-appro}
  n_{\rm e}(R,z)=0.04[1+(R/R_{c})^2+(z/z_{c})^2]^{-3\beta/2} ~~\text{cm}^{-3},
\end{equation}
where $R_{c}=0.58$ kpc and $z_{c}=0.45$ kpc. Due to significant X-ray absorptions toward the GC, the inner core size ($R_{c}$, $z_{c}$) can not be well constrained by our study, and the $\beta$ models with $R_{c} \approx z_{c} \lesssim 0.5$ kpc are all acceptable. Ignoring the potentially very small inner core, our best-fit CGM model for the inner Galaxy (Eq. \ref{eq-beta-flat-appro}) can be simply rewritten as a power law in radius $n_{\rm e}(r)=0.01(r/1 \text{~kpc})^{-1.5}$ cm$^{-3}$. { The slope of the initial CGM density profile is a key parameter affecting the morphology and the X-ray surface brightness distribution of the Fermi bubbles.}

We rule out the relatively flat MB and G20 models for the CGM distribution in the inner Galaxy. The G20 model \citep{Guo2020} could match the best-fit power-law distribution ($n_{\rm e} \propto r^{-1.5}$) when properly choosing the model parameters $\alpha=1.5$ and $r_{1} \lesssim 0.5$ kpc. The centrally cuspy NFW density model is not steep enough to be consistent with observations, either. While we could not rule out all disk-like models, both the L17 and S19 models are inconsistent with some observations. The L17 model has a very large density core which leads to cylindrical Fermi bubbles with very wide bases. The S19 model has a very small core, but beyond the core, the gas density drops exponentially and reaches $<10^{-5}$ cm $^{-3} $ at $|z|> 3$ kpc, which is too low to explain the X-ray observations.

In Fig. \ref{fig-den1d-cali}, we extend our constrained density models to large radii and find that the extrapolated gas densities at $r\sim 50-90$ kpc are appreciably lower than the recent halo gas density estimates from the ram-pressure stripping calculations by \citet{Salem2015} and \citet{Gatto2013}. This suggests that the CGM density profile either flattens out or has one or more discontinuities at large radii. The potential density discontinuities may be related to the recently discovered eROSITA bubbles located outside the Fermi bubbles \citep{predehl20, Nakahira2020}. 
{ In our jet-shock model, the propagating forward shock driven by the GC AGN jet event occurring about 5 Myr ago corresponds to the Fermi bubble edge at the current time, and thus could not explain the origin of the eROSITA bubbles. In this model, the eROSITA and Fermi bubbles are two unrelated phenomena, and the former event occurred earlier. In this case, the derived CGM density profile $n_{\rm e}(r)=0.01(r/1 \text{~kpc})^{-1.5}$ cm$^{-3}$ in our study may correspond to the hot gas distribution in the inner 10-kpc region of the eROSITA bubbles about $5$ Myr ago.}

In our simulations, we have assumed that the initial halo gas has a uniform temperature $T=0.2$ keV and a uniform metallicity $Z=0.4 Z_{\odot}$. If $T$ or $Z$ is higher, the normalization of the constrained CGM density profile should be lower to fit the observed X-ray surface brightnesses of the Fermi bubbles. If $T$ or $Z$ decreases with radius, the initial CGM density profile should be relatively flatter than in our best-fit model, which may affect the morphology of the Fermi bubbles. At $T=0.2$ keV, the best-fit Z20 density distribution is under hydrostatic equilibrium in the Galactic potential. If $T$ is higher or the size of the inner density core is much smaller than $0.5$ kpc, the initial halo gas in the inner Galaxy is expected to be in an outflowing state.

\acknowledgments
{ We thank the anonymous referee for an insightful report. We thank Yuning Zhang and Zhen Yan for helpful discussions. This work was supported partially by the National Natural Science Foundation of China (Nos. 11873072 and 11633006), the Natural Science Foundation of Shanghai (No. 18ZR1447100), and the Chinese Academy of Sciences through the Key Research Program of Frontier Sciences (Nos. QYZDB-SSW-SYS033 and QYZDJ-SSW-SYS008). FG has benefitted from some relevant talks in a KITP workshop, supported in part by the National Science Foundation under Grant No. NSF PHY-1748958. The simulations were performed with the high performance computing resources in the Core Facility for Advanced Research Computing at Shanghai Astronomical Observatory.}

\bibliography{ms}

\end{document}